\documentclass[structabstract]{aa}  
\usepackage{natbib}
\usepackage{graphicx}
\usepackage{lscape}
\usepackage{longtable}
\usepackage{txfonts}
\usepackage{color}                                                                           

\begin{document}

   \title{Metallicity of solar-type stars with debris discs and planets
    \thanks{
     Based on observations collected at the Centro Astron\'omico Hispano
     Alem\'an (CAHA) at Calar Alto, operated jointly by the Max-Planck Institut
     f\"{u}r Astronomie and the Instituto de Astrof\'isica de Andaluc\'ia (CSIC);
     observations made with the Italian Telescopio Nazionale Galileo (TNG) operated
     on the island of La Palma by the Fundaci\'on Galileo Galilei of the INAF
     (Istituto Nazionale di Astrofisica); 
     observations made with the Nordic Optical Telescope, operated
     on the island of La Palma jointly by Denmark, Finland, Iceland,
     Norway, and Sweden, in the Spanish Observatorio del Roque de los
     Muchachos of the Instituto de Astrofisica de Canarias; and
     data obtained from the ESO Science Archive Facility.} \fnmsep
     \thanks{Tables 1 and 4 are only available in the electronic version of the paper or
     at the CDS via anonymous ftp to cdsarc.u-strasbg.fr (130.79.128.5)
     or via http://cdsweb.u-strasbg.fr/cgi-bin/qcat?J/A+A/
     }}


   \author{J. Maldonado 
          \inst{1}
          \and  C. Eiroa
          \inst{1}
          \and  E. Villaver 
          \inst{1}  
          \and  B. Montesinos
          \inst{2}
          \and  A. Mora
          \inst{3}
          }

        \institute{Universidad Aut\'onoma de Madrid, Dpto. F\'isica Te\'orica, M\'odulo 15,
             Facultad de Ciencias, Campus de Cantoblanco, E-28049 Madrid, Spain,
             \and
             Centro de Astrobiolog\'ia (INTA-CSIC), LAEFF Campus, European Space Astronomy Center (ESAC),
             P.O. Box 78, E-28691 Villanueva de la Ca{\~n}ada, Madrid, Spain
             \and
             ESA-ESAC Gaia SOC. P.O. Box 78, E-28691 Villanueva de la Ca{\~n}ada, Madrid, Spain 
             }

   \offprints{J. Maldonado, \\  \email{jesus.maldonado@uam.es}}
   \date{Received 10 January 2012; accepted 06 February 2012}

 
  \abstract
   {Around  16\% of  the solar-like  stars  in our  neighbourhood show
   IR-excesses due  to  dusty debris  discs and a fraction of them are known to host planets.
   Determining whether these  stars follow any  special trend in their  properties is
  important  to understand  debris disc and planet formation.
   }
   {We aim to determine in a homogeneous way the metallicity of a sample of stars with
    known debris discs and planets. We attempt to identify trends related to
    debris discs and planets around solar-type stars.
    }
   {Our analysis includes the calculation of the fundamental
    stellar parameters $T_{\rm eff}$, $\log g$, microturbulent velocity, and metallicity
    by applying the iron ionisation
    equilibrium conditions to several isolated Fe~{\sc i} and Fe~{\sc ii} lines.
    High-resolution \'echelle spectra ($R \sim 57000$) from 2-3 meter class telescopes are used.
    Our derived metallicities are compared with other results in the literature, which
    finally allows us to extend the stellar samples in a consistent way. 
   }
   {The  metallicity  distributions  of  the different  stellar  samples
   suggest that there is a transition toward higher metallicities from stars with neither debris discs nor planets
   to stars  hosting  giant planets. Stars with debris discs and stars with neither debris nor planets
   follow a similar metallicity distribution, although the distribution of the first
   ones  might be  shifted towards  higher metallicities. Stars with
   debris  discs and planets have the  same metallicity  behaviour as
   stars hosting planets, irrespective  of whether the planets are low-mass or
   gas  giants.  In the  case of  debris discs  and giant  planets, the
   planets are usually  cool, - semimajor axis larger  than 0.1 AU (20
   out of 22 planets), even $\approx$ 65\% have semimajor axis larger
   than 0.5 AU. The data also suggest that stars with debris discs
   and cool giant  planets tend to have a low  dust luminosity, 
   and  are   among  the  less   luminous  debris  discs   known.
   We also find evidence of an
   anticorrelation between  the luminosity of the dust  and the planet
   eccentricity. 
   }
    {Our data show that the presence of planets, not the debris
    disc, correlates with the stellar metallicity. The results confirm
    that core-accretion models represent suitable scenarios for debris disc and
    planet formation. These conclusions are based on a
    number of stars with discs and planets considerably larger than in previous
    works, in particular stars hosting low-mass planets and debris discs.
    Dynamical instabilities produced by eccentric giant planets could explain the
    suggested dust luminosity trends observed for stars with debris discs and planets.
    }
   \keywords{techniques: spectroscopic - stars: abundances - stars: circumstellar matter -stars: late-type -stars: planetary systems}

   \maketitle

\section{Introduction}

 Understanding the origin and evolution of planetary systems is one of
 the major goals of modern  astrophysics. The unexpected
 discovery  by the {\it IRAS} satellite of infrared excesses around main-sequence
 stars \citep{1984ApJ...278L..23A}
 was attributed to the presence of  faint dusty discs, produced
 by collisional events within  a significant population of invisible
 left-over planetesimals. The discovery of
 these  so-called \textit{debris discs} demonstrated
 that  planetesimals  are
 more  common  than had been  previously  thought,  revealing that  the
 initial    steps    of    planetary    formation    are    ubiquitous
 \citep[e.g.][]{1993prpl.conf.1253B}.  This   realisation  has  been
 complemented in the past 15 years with the detection of more
 than 700 exoplanets  orbiting stars other than the Sun
 \footnote{http://exoplanet.eu/}.

 More recent studies have found that more than 50\% of solar-type stars harbor at least
 one planet of any mass with a period of up to 100 days, and about 14\% of this
 type of stars have planetary companions more massive than 50 M$_{\oplus}$
 with periods shorter than 10 years \citep{2011arXiv1109.2497M}.
 It is well-established that the percentage
 of stars hosting gas-giant planets increases with the metal content, up
 to 25\% for stars with metallicities higher than  +0.30 dex 
 \citep[e.g.][]{2004A&A...415.1153S,2005ApJ...622.1102F}.
 On the other hand, stars that host less massive planets, Neptune-like or super Earth-like 
 planets (M$_{\rm p}$ $<$ 30 M$_{\oplus}$), do not tend to be metal-rich
 \citep[][and references therein]{2010ApJ...720.1290G,2011arXiv1109.2497M,2011A&A...533A.141S}.
 In terms of the metallicity of evolved stars
 (late-type subgiants and red giants)
 hosting planets, previous results have been based on the analysis of small and inhomogeneous
 samples that even produce contradictory results,
 while
 these stars are metal-poor in the cases of
 \cite{2007A&A...473..979P} and \cite{2010ApJ...725..721G}, 
 they show  metal enrichment according to \cite{2007A&A...475.1003H}.

 Debris discs are, strictly speaking, signatures of planetesimal systems.
 About 16\% of the 
 main-sequence solar-like (spectral types  F5-K5) stars are known to show  an excess at
 70 $\mu$m \citep[e.g.][]{Trilling08}. If planetesimals were the building
 blocks of planets and, at the same time, the raw material from
 which debris discs form, their host stars might be expected to 
 have similar properties. However, the incidence of debris discs is no higher
 around planet-host stars than around stars without detected planets
 \citep{2009ApJ...700L..73K}, and several works do not find any correlation
 between the presence of a debris disc and the metallicity, or any other
 characteristic, of the stars with planets 
 \citep[e.g.][]{2005ApJ...622.1160B,2006A&A...452..921C,2006MNRAS.366..283G,
 2007ApJ...658.1312M,2009ApJ...705.1226B,2009ApJ...700L..73K}.

 In this paper, we revisit the analysis of the properties of solar-type
 stars hosting planets and/or debris discs. One of the motivations
 is the increase with respect to previous works of $\sim$ 50\%, in the number of 
 stars with known debris discs and planets, in particular those
 associated with low-mass planets (M$_{\rm p}$ $\lesssim$ 30M$_{\oplus}$). 
 We distinguish three different
 categories: stars with known debris discs but no planets (SWDs hereafter), 
 stars with known debris discs and planets (SWDPs), and stars with known planets but no
 discs (SWPs). In addition, we consider a comparison sample of stars with no
 detected planets and no detected debris discs (SWODs). 
 We use our own high-resolution \'echelle spectra to homogeneously
 determine some of the stellar properties, particularly metallicity,
 and in a second step we compare our spectroscopic results 
 with published results. This allows us to increase coherently the stellar samples
 analysed in this work.

\section{Observations} 

\subsection{The stellar sample}
\label{stellar_sample}

 A list of stars
 with known debris discs, SWDs, 
 was compiled by carefully checking the
 works of \cite{Habing01},
 \cite{Spangler01}, \cite{2005ApJ...634.1372C},
 \cite{Beichman06}, \cite{Bryden06}, \cite{Moor06}, \cite{2006ApJ...644L.125S},
 \cite{2007ApJ...658.1312M}, \cite{Rhee07}, \cite{2007ApJ...658.1289T},
 \cite{Trilling08}, \cite{2009ApJ...705.1226B}, \cite{2009ApJ...700L..73K}, 
 \cite{0004-637X-698-2-1068}, \cite{Tanner2009ApJ}, \cite{2010ApJ...710L..26K},
 \cite{DodsonRobinson11}, and \cite{2011ApJS..193....4M}.
 These debris discs were discovered by the {\it IRAS}, {\it ISO}, 
 and {\it Spitzer} telescopes.
 We compiled a total list of 305 stars, from which we retained for study only
 the solar-type stars (Hipparcos spectral type
 between F5 and K2-K3), leading to a total of 136 stars. Most of the debris discs around these stars were
 detected at {\it Spitzer}-MIPS 70 $\mu$m, with fractional dust luminosities
 of the order of $10^{-5}$ and higher \citep{Trilling08}.

 To build the {\it comparison} sample of stars without discs (SWODs), we selected
 from the aforementioned works stars in which IR-excesses were not found
 at 24 and 70 $\mu m$ by
 {\it Spitzer}.
 As before, only solar-type
 stars were considered, leading to 150 stars. Since {\it Spitzer} is limited up to fractional luminosities of
 L$_{\rm dust}$/L$_{\star}$ $\gtrsim$ 10$^{-5}$, we cannot rule out the possibility
 that some of these stars have fainter discs. 

 To avoid the effects of planets,
 planet-hosting stars in both the SWD and SWOD samples were removed, after checking the
 Extrasolar Planets Encyclopedia\footnote{http://exoplanet.eu/}. 
 The final number of stars in the SWDs sample is 107: 49 F-type stars, 37 G-type stars,
 and 21 K-type stars. The SWODs sample contains 145 stars: 62 F-type stars, 65
 G-type stars, and 18 K-type stars.
 Table~\ref{sample_stars} lists the stars in the SWD and SWOD samples, their
 properties, and  references to debris disc detection.


\begin{table}
\centering
\caption{The SWD and SWOD samples. Only the first eight lines of the SWD sample and the references are presented here;
the full version of the table is available as online material.}
\label{sample_stars}
\begin{scriptsize}
\begin{tabular}{llllllrl}
\hline\noalign{\smallskip}
HIP & HD & SpType & V & distance & $\log$(Age) & [Fe/H]$^{\dag}$  & Ref \\
    &    &        &   & (pc)     &   (yr)      &  (dex)           &     \\
\hline\noalign{\smallskip}
\multicolumn{8}{c}{Stars with known debris discs.}\\
\hline\noalign{\smallskip}
171     &       224930  &       G3V     &       5.80    &       12.17   &       9.60    &       -0.72  (a)      &       13      \\
490     &       105     &       G0V     &       7.51    &       39.39   &       8.34    &       -0.03  (b)      &        2      \\
544     &       166     &       K0V     &       6.07    &       13.67   &       9.16    &       0.15   (a)      &        5      \\
682     &       377     &       G2V     &       7.59    &       39.08   &       8.34    &       0.12   (b)      &        6      \\
1481    &       1466    &       F8/G0   &       7.46    &       41.55   &       8.34    &       -0.22  (c)      &        7      \\
1598    &       1562    &       G0      &       6.97    &       24.80   &       9.79    &       -0.32  (a)      &       13      \\
1599    &       1581    &       F9V     &       4.23    &       8.59    &       9.58    &       -0.29  (a)      &        9      \\
2843    &       3296    &       F5      &       6.72    &       45.05   &               &        0.02  (c)      &        9      \\
\hline\noalign{\smallskip}
\multicolumn{8}{l}{$^{\dag}$ (a) This work; (b)  \cite{2005ApJS..159..141V}; (c) \cite{2004A&A...418..989N} }\\
\multicolumn{8}{l}{(b) and (c) values are set into our metallicity scale as described in Section~\ref{previous_work}.}\\
\hline
\multicolumn{8}{l}{(1) \cite{Habing01}; (2)\cite{Spangler01}; (3) \cite{2005ApJ...634.1372C};} \\
\multicolumn{8}{l}{(4) \cite{Beichman06}; (5) \cite{Bryden06}; (6) \cite{Moor06};}\\
\multicolumn{8}{l}{(7) \cite{2006ApJ...644L.125S}; (8) \cite{Rhee07}; (9) \cite{Trilling08};  }\\
\multicolumn{8}{l}{(10) \cite{2009ApJ...700L..73K}; (11) \cite{0004-637X-698-2-1068}; (12) \cite{Tanner2009ApJ};}\\
\multicolumn{8}{l}{(13) \cite{2010ApJ...710L..26K}; (14) \cite{2011ApJS..193....4M};}\\
\noalign{\smallskip}\hline\noalign{\smallskip}
\end{tabular}
\end{scriptsize}
\end{table}

\subsection{Possible biases}
\label{possible_biases}

\begin{figure}
\centering
\includegraphics[angle=270,scale=0.45]{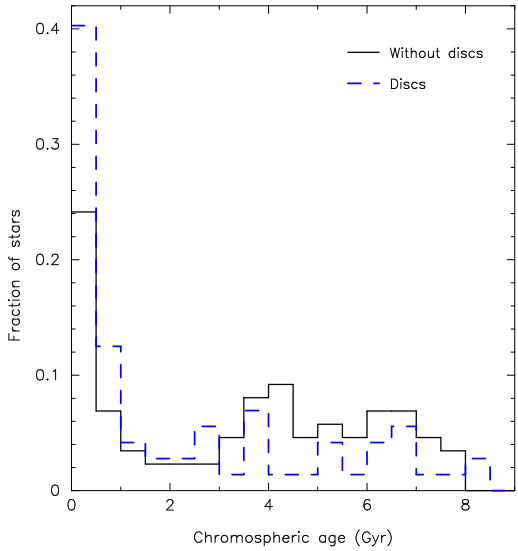}
\caption{Age distribution for stars in the SWOD (continuous-black line) and the
 SWD (dotted-blue line) samples.}
\label{histogram_ages}
\end{figure}

 Metallicity reflects the enrichment history of the 
 ISM \citep[see e.g.][]{1995ApJS...98..617T}. It is, therefore, important
 to determine whether the SWD and SWOD
 samples have randomly selected stellar hosts in terms of age
 and distance, which are
 the parameters
 most likely to affect the metal
 content of a star. 
 In this respect, we compared the distances and stellar ages of both samples;
 the results are given in Table~\ref{bias}. 
 Distances are from the updated Hipparcos parallaxes
 \citep{Leeuwen} and stellar ages were computed from the $\log$ R'$_{\rm HK}$ values given
 by \cite{2010A&A...521A..12M}
 or from the literature if no value was available in that work. The relationship provided by
 \citet[][Eq. 3]{2008ApJ...687.1264M}
 was used to compute the ages. This relationship has an accuracy of 15-20\%
 for young stars, i.e. younger than 0.5 Gyr, and at older age, uncertainties
 can grow by up to 60\%. 
 The age distributions 
 are shown in Figure~\ref{histogram_ages}.


\begin{table}
\centering
\caption{Comparison between the properties of the SWDs and the SWODs samples.}
\label{bias}
\begin{scriptsize}
\begin{tabular}{lcccccc}
\hline\noalign{\smallskip}
               &  \multicolumn{3}{c}{\textbf{SWDs}} & \multicolumn{3}{c}{\textbf{SWODs }} \\
               &  \multicolumn{3}{c}{\hrulefill}    & \multicolumn{3}{c}{\hrulefill}      \\
               & Range      & Mean        &  Median &   Range       & Mean    & Median    \\
\hline\noalign{\smallskip}
 Distance (pc)    & 3.6/134 & 32.0 & 24.6 & 5.8/53  & 24.1 & 20.6  \\
 $\log$[Age (yr)] & 7.2/9.9 & 9.0  & 9.0  & 7.6/9.9 & 9.2  &  9.6  \\
\hline\noalign{\smallskip}
 SpType (\%)      & \multicolumn{3}{l}{45.8 (F); 34.6 (G); 19.6 (K)} & \multicolumn{3}{l}{42.8 (F); 44.8 (G); 12.4 (K)} \\
\noalign{\smallskip}\hline\noalign{\smallskip}
\end{tabular}
\end{scriptsize}
\end{table}


\begin{figure}
\centering
\includegraphics[angle=270,scale=0.45]{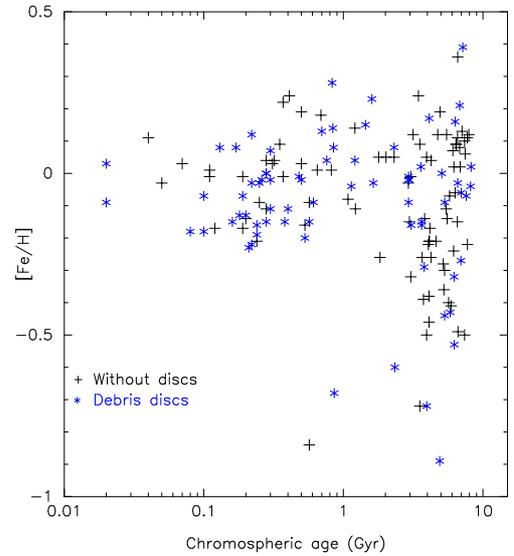}
\caption{[Fe/H] versus age for the stars in the SWOD (black crosses) and in the
SWD (blue asterisks) samples.}
\label{metallicity_ages}
\end{figure}

 We found a difference between the two samples
 in terms of the age, with the SWDs containing 15\% more  stars
 younger than 500 Myr. Type II and Type Ia supernova (SNe) are the two
 sources of Fe production, each operating on
 different timescales and accounting for very different amounts of 
 the total Fe injected into the ISM. While Type Ia SNe are
 the major producers of Fe in galaxies \citep[see e.g.][]{1986A&A...154..279M},
 their injection timescales are, according to the most recent estimates, longer
 than 1 Gyr \citep{2009A&A...501..531M}. In the solar neighbourhood, this
 1 Gyr timescale, although uncertain, is the time at which the Fe production
 from SNe Ia starts to become important \citep{2001ApJ...558..351M}.
 On the other hand, Type II SNe are expected to account for only 30\%
 of the total yield of Fe \citep{1986A&A...154..279M} but 
 are expected to do so on a shorter timescale (3-5 Myr). A high rate of local SN type II explosions has been
 estimated to  explain the local bubble \citep{2001ApJ...560L..83M}, namely
 20 SN type II explosions within 150 pc of the Sun in the past 11 Myr \citep{2002PhRvL..88h1101B}.
 The youngest stars in the SWD and SWOD samples have ages of 15 Myr and 35 Myr, respectively, with
 a larger number of young SWDs in the first 500 Myr bin. The paucity of SNe type II
 in the Galaxy (typical rate of $\approx$ 1 SNe Myr$^{\rm -1}$) and all the stars being
 at relatively close distances from the Sun (less than 130 pc) make it very unlikely
 that the two samples have experienced different enrichment histories. 
 We have, however, explored this possibility in Figure~\ref{metallicity_ages},
 where we plot the metallicity versus age (Section~\ref{seccion_resultados}) 
 for the two samples. As we can see, the SWDs and SWODs have similar
 behaviours. Young stars in the SWDs sample do not seem to have higher metallicities,
 so we can rule out a possible chemical evolution in the SWD sample. 


 We have also checked whether there is a difference between the SWD and
 SWOD samples in terms of distance that might affect their metallicity
 distributions. After all, the SWD sample  contains stars out to a larger volume  
 than that of the SWODs and could possibly include stars with
 a different chemical evolution. \cite{2000ApJ...532.1192G} studied
 the scatter in the age-metallicity relation for F and G dwarf stars
 in the solar neighbourhood up to 80 pc, and found that their stars at distances 30-80 pc from
 the Sun are more metal-poor than those within 30 pc.  \cite{2000ApJ...532.1192G}
 attributed this difference to the possible consequence of a selection bias
 in the analysed sample. 

 We certainly cover the same volume of stars in our homogeneous SWD and
 SWOD subsamples (see Section~\ref{homogeneous_section}),  
 since they are located within 25 pc of the Sun \citep{2010A&A...521A..12M}.
 We do not find any chemical distinction between these two subsamples.
 If in the full sample (see Section~\ref{full_sample})
 we had
 a selection bias due to the larger distance of the SWODs, we would
 expect its metallicity distribution to show a larger dispersion owing to
 a possible contamination by stars not born in the solar neighbourhood. 
 We have the opposite case, 
 where the full samples of both SWODs and SWDs have
 a smaller dispersion than the volume-limited homogeneous
 SWDs and SWODs subsamples (see Figure~\ref{histogram}).

 In short, we believe that we have a randomly selected sample of stars 
 in terms of their chemical history, although the SWD and SWOD samples
 show some differences in age and distance.

 \subsection{Spectroscopic observations}
 \label{spectroscopic_observations}

 The high-resolution spectra used in this work are the same as in 
 \cite{2010A&A...521A..12M}, where a complete description of
 the observing runs and the reduction procedure can be found. 
 In brief, the data were taken with the following spectrographs and
 telescopes:
 i) FOCES \citep{foces}  at the  2.2 meter telescope of the Calar Alto
 observatory (Almer\'ia, Spain);
 ii) SARG \citep{SARG} at the TNG, La Palma (Canary Islands,  Spain);
 and iii) FIES \citep{1999anot.conf...71F} at the NOT, La Palma.
 We also used additional spectra  from  the  public
 library  ``S$^{4}$N'' \citep{s4n}, which  contains spectra  taken with
 the 2dcoud\'{e}  spectrograph at McDonald Observatory and the FEROS
 instrument at  the ESO  1.52 m  telescope in La  Silla, and  from the
 ESO/ST-ECF  Science Archive Facility  \footnote{http://archive.eso.org/cms/}
 (specifically FEROS spectra). 
 Table~\ref{tabla_espectrografos} lists the spectral range
 and resolving power of each of the spectrographs.
 The number of stars covered by these spectra are 35 (33\%) and 58 (40\%) for the
 SWD and SWOD samples, respectively. Thus, we consider additional data from the literature
 to analyse the whole set of stars in
 both samples (see Section~\ref{full_sample}).


\begin{table}
\centering
\caption{
Properties of the different spectrographs used in this work.
}
\label{tabla_espectrografos}
\begin{tabular}{lcc}
\hline\noalign{\smallskip}
Spectrograph &  Spectral range (\AA)  & Resolving power  \\
\hline 
FOCES        &  3470-10700            &  57000           \\
SARG         &  5500-10100            &  57000           \\
FIES         &  3640-7360             &  67000           \\
FEROS        &  3500-9200             &  42000           \\
Mc Donald    &  3400-10900            &  60000           \\
\hline
\end{tabular}
\end{table}

 \subsection{Analysis}
 \label{analysis}

 The
 stellar  parameters  $T_{\rm eff}$,  $\log g$,  microturbulent  velocity
 ($\xi_{t}$), and [Fe/H], are determined using the code TGV developed
 by \cite{2002PASJ...54..451T},  which is based on iron-ionisation
 equilibrium  conditions, a methodology that is widely applied
 to solar-like stars (spectral types F5/K2-K3).
 Iron abundances  are  computed  for a  well-defined
 set of {\rm Fe~{I}}  and {\rm Fe~{II}} lines.  Basically, the
 stellar  parameters are  adjusted until:  i) no  dependence  is found
 between the abundances derived from  {\rm Fe~{I}} lines and the lower
 excitation potential of the lines; ii) no dependence is found between
 the abundances derived from the  {\rm Fe~{I}} lines and their equivalent
 widths;  and  iii) the derived   mean  {\rm  Fe~{I}}  and  {\rm  Fe~{II}}
 abundances are the same.     We   list    the   lines    used   in
 Table~\ref{feI_feII_lines}.  
 We  are  aware  that  ideally  all  our
 targets  should have  been observed  with the  same  spectrograph 
 using the same configuration. Nevertheless, all the spectra used here
 have a similar resolution, and cover enough {\rm Fe} lines to provide
 a high-quality metallicity determination.  Only for  the SARG
 spectra is the number of {\rm  Fe~{II}} lines slightly lower (6 out
 of 13, begining in the 6432.68 \space \AA \space line).


\begin{table}
\centering
\caption{{\rm Fe~{I}} and {\rm Fe~{II}} lines used to compute abundances. Wavelengths
are given in Angstroms (\AA).}
\label{feI_feII_lines}
\begin{tabular}{llll}
\hline\noalign{\smallskip}
\multicolumn{3}{c}{\textbf{\rm Fe~{I}}} & {\textbf{\rm Fe~{II}}} \\
\multicolumn{3}{c}{\hrulefill}          & \hrulefill             \\
4389.25	&	6173.34	&	6699.14	&	4576.34	 \\
4445.48	&	6180.21	&	6739.52	&	4620.52	 \\
5225.53	&	6200.32	&	6750.16	&	4656.98	 \\
5247.06	&	6219.29	&	6752.71	&	5234.63	 \\
5250.22	&	6232.65	&	6793.27	&	5264.79	 \\
5326.15	&	6240.65	&	6804.00	&	5414.08	 \\
5412.79	&	6265.14	&	6804.28	&	5525.13	 \\
5491.84	&	6271.28	&	6837.01	&	6432.68	 \\
5600.23	&	6280.62	&	6854.83	&	6516.08	 \\
5661.35	&	6297.80	&	6945.21	&	7222.40	 \\
5696.09	&	6311.51	&	6971.94	&	7224.46	 \\
5701.55	&	6322.69	&	6978.86	&	7515.84	 \\
5705.47	&	6353.84	&	7112.17	&	7711.73	 \\
5778.46	&	6481.88	&	7401.69	&		 \\
5784.66	&	6498.95	&	7723.21	&		 \\
5855.08	&	6518.37	&	7912.87	&		 \\
5909.98	&	6574.23	&	8075.16	&		 \\
5956.70	&	6581.21	&	8204.11	&		 \\
6082.72	&	6593.88	&	8293.52	&		 \\
6120.26	&	6609.12	&	8365.64	&		 \\
6137.00	&	6625.03	&		&		 \\
6151.62	&	6667.72	&		&		 \\
\noalign{\smallskip}\hline\noalign{\smallskip}
\end{tabular}
\end{table}


\begin{figure*}
\centering
\includegraphics[angle=270,scale=0.45]{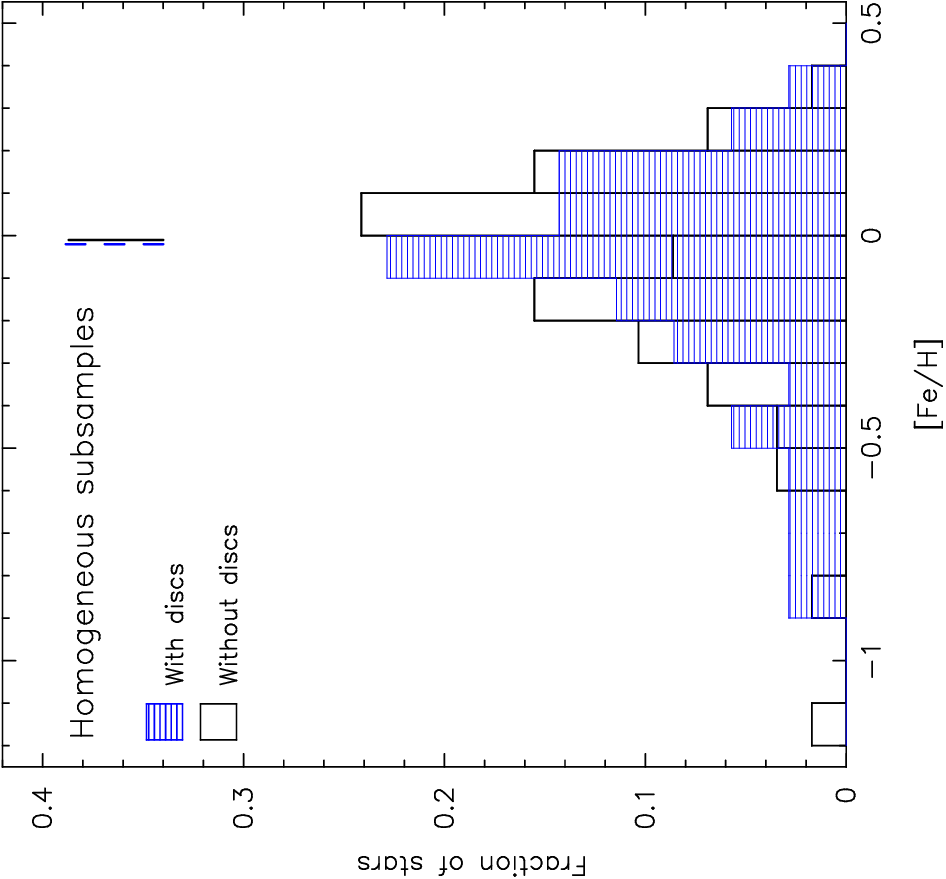}
\includegraphics[angle=270,scale=0.45]{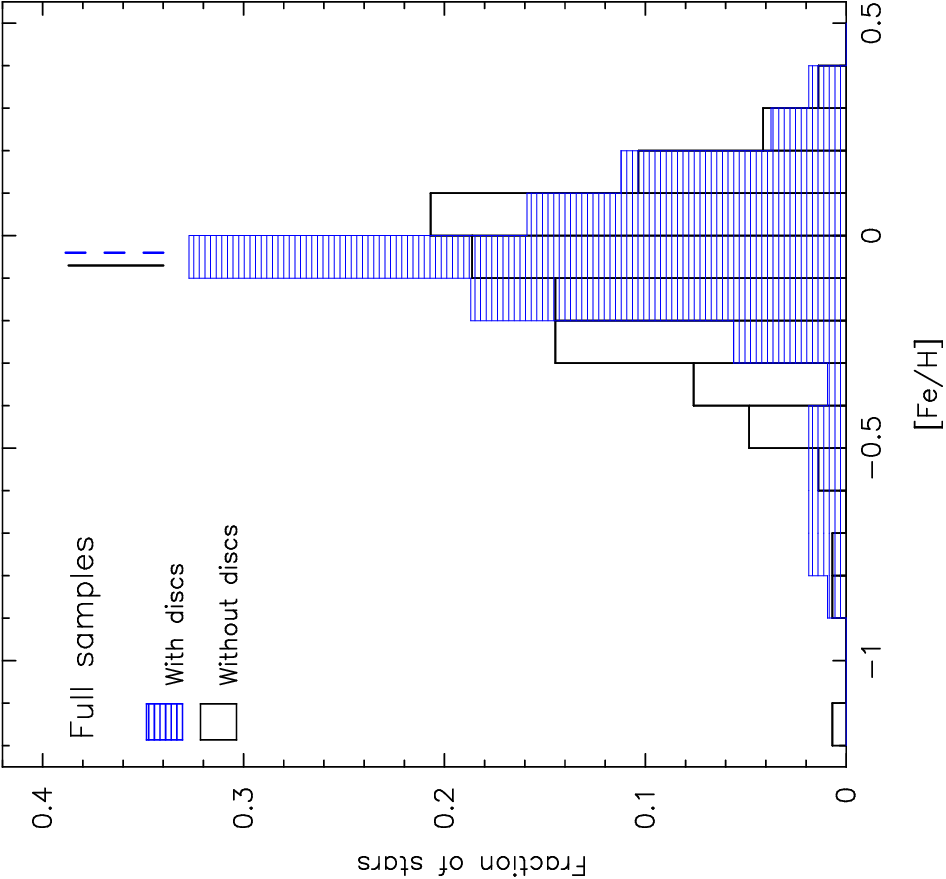}
\caption{Normalized  metallicity  distribution  of the  stars  without
  debris  discs (SWODs, empty  histogram), and  the stars  with debris
  discs (SWDs, blue  histogram shaded at 0 degrees).  Median values of
  the  distributions  are  shown  with  vertical  lines.  Left  panel:
  distributions  of the stars in the  homogeneous sample, i.e.,  metallicities
  computed from our own spectra. Right  panel: distributions of the full
  stellar sample (see text).} 
\label{histogram}
\end{figure*}

  Equivalent  widths are  obtained by fitting  the lines  with  a Gaussian
  profile using  the IRAF \footnote{IRAF is distributed  by the National
  Optical Astronomy Observatory, which  is operated by the Association
  of Universities for Research in Astronomy, Inc., under contract with
  the National Science Foundation.} task \textit{splot}.  Stars with
  significant  rotational velocities, $v \sin i$, have lines affected  by blending,
  complicating the application of this method. Stars with
  $v \sin i$ $\gtrsim$ 15-20 kms$^{-1}$ typically do not have enough isolated lines
  to  obtain  accurate  parameters.  This  has a  small  impact  on  our
  estimates since  we consider  stars with spectral types  F5 or
  later,  with  typical v$\sin$i  values  in  the  range 3-9  kms$^{-1}$
  \citep{2010A&A...520A..79M}. The estimated stellar parameters and iron
  abundances are  given in Table~\ref{parameters_table}.


\begin{table*}
\centering
\caption{
 Estimated physical parameters with uncertainties
 for the stars measured in this work. Columns 8 an 10
 give the mean iron abundance derived from Fe~{\sc I} and Fe~{\sc  II} lines,
 respectively, while columns 9 and 11 give the corresponding number of lines.
 The rest of the columns are self-explanatory. Only the first eight lines are
 shown here; the full version of the table is available online.
 }
\label{parameters_table}
\begin{tabular}{lllcccccccccl}
\hline\noalign{\smallskip}
HIP &  HD &  SpType & T$_{\rm eff}$ &  $\log g$      & $\xi_{t}$     & [Fe/H] &  $\left\langle A({\rm Fe}~{\rm I}) \right\rangle$ & n$_{\rm I}$ &
 $\left\langle A({\rm Fe}~{\rm II}) \right\rangle$  & n$_{\rm II}$ &  Spec.$^{\dag}$ \\
    &     &     & (K)   & (${\rm cm s^{-2}}$)  & (${\rm km s^{-1}}$) & dex  &      &      &      &      &      \\
 (1)& (2) & (3) & (4)   & (5)            & (6)           & (7)  & (8)  & (9)  & (10) & (11) & (12) \\
\hline\noalign{\smallskip}
\multicolumn{12}{c}{Stars with known debris discs.}\\
\hline\noalign{\smallskip}
171	&	224930	&	G3V	&	5491	$\pm$	31	&	4.75	$\pm$	0.12	&	0.92	$\pm$	0.40	&	-0.72	$\pm$	0.08	&	6.78	$\pm$	0.11	&	52	&	6.78	$\pm$	0.12	&	12	&	4	\\
544	&	166	&	K0V	&	5575	$\pm$	51	&	4.68	$\pm$	0.14	&	1.05	$\pm$	0.25	&	0.15	$\pm$	0.05	&	7.65	$\pm$	0.06	&	57	&	7.65	$\pm$	0.06	&	13	&	4	\\
1598	&	1562	&	G0	&	5603	$\pm$	36	&	4.30	$\pm$	0.12	&	0.67	$\pm$	0.27	&	-0.32	$\pm$	0.06	&	7.18	$\pm$	0.07	&	58	&	7.18	$\pm$	0.09	&	12	&	1	\\
1599	&	1581	&	F9V	&	5809	$\pm$	39	&	4.24	$\pm$	0.12	&	1.30	$\pm$	0.30	&	-0.29	$\pm$	0.06	&	7.21	$\pm$	0.08	&	59	&	7.22	$\pm$	0.10	&	13	&	5	\\
5336	&	6582	&	G5V	&	5291	$\pm$	32	&	4.57	$\pm$	0.11	&	0.82	$\pm$	0.42	&	-0.89	$\pm$	0.08	&	6.61	$\pm$	0.11	&	55	&	6.62	$\pm$	0.12	&	12	&	4	\\
5862	&	7570	&	F8V	&	6111	$\pm$	35	&	4.42	$\pm$	0.10	&	1.35	$\pm$	0.17	&	0.17	$\pm$	0.03	&	7.67	$\pm$	0.04	&	50	&	7.67	$\pm$	0.04	&	13	&	6	\\
5944	&	7590	&	G0	&	5951	$\pm$	39	&	4.65	$\pm$	0.11	&	1.04	$\pm$	0.38	&	-0.02	$\pm$	0.05	&	7.48	$\pm$	0.06	&	48	&	7.48	$\pm$	0.08	&	11	&	1	\\
7576	&	10008	&	G5	&	5293	$\pm$	68	&	4.90	$\pm$	0.19	&	0.39	$\pm$	0.45	&	0.08	$\pm$	0.06	&	7.58	$\pm$	0.07	&	53	&	7.58	$\pm$	0.10	&	9	&	1	\\
\noalign{\smallskip}\hline\noalign{\smallskip}
\multicolumn{12}{l}{$^{\dag}$Spectrograph: {\bf(1)} CAHA/FOCES; {\bf(2)} TNG/SARG; {\bf(3)} NOT/FIES; {\bf(4)} S$^{4}$N-McD;
{\bf(5)} S$^{4}$N-FEROS; {\bf(6)} ESO/FEROS}\\ 
\end{tabular}
\end{table*}

\section{Results}
\label{seccion_resultados}

 \subsection{Homogeneous analysis}
 \label{homogeneous_section}

 In a first step, we consider the 35 SWDs and 58 SWODs whose
 metallicities were estimated directly in this work. 
 The stars in these {\it homogeneous samples} are listed in
 Table~\ref{parameters_table}, and are marked in Column 7 of Table~\ref{sample_stars}
 as well. 
 Figure~\ref{histogram} (left panel) shows the normalized distribution of these
 stars. Both distributions are very similar.   
 The metallicity distribution of the SWOD sample spreads over a large range
 containing both metal-poor and metal-rich stars, from -1.12 to 0.36 dex. The mean
 metallicity of the distribution is -0.09 dex with an RMS dispersion of 0.27 dex.
 The distribution of the SWDs spans a slightly narrower range, from -0.89 to 0.35 dex, with 
 a mean value of -0.10 dex and a dispersion of 0.28 dex. 
 Since the mean of a distribution is strongly affected by the presence of outliers,
 we consider the median 
 as a more representative value.
 The median values for the SWOD and SWD distributions are -0.01 and -0.02,
 respectively.
 To assess whether both distributions are equal from a statistical point of
 view, a two-sample Kolmogorov-Smirnov (K-S) test was performed 
 (details about how the K-S test is applied in this paper
 are given in Appendix~\ref{apendice}).
 The maximum difference
 between the SWD and SWOD cumulative distribution functions is only $\sim$ 0.11,
 while the likelihood that both samples have the same parent distribution is around 94\%.

\subsection{Comparison with previous works}
\label{previous_work}

 The spectroscopic observations performed by \cite{2010A&A...521A..12M} were limited to 25 pc in distance
 and, therefore, do not cover all SWDs and SWODs in Table~\ref{sample_stars}.
 Thus, we use the data  of \citet[][NO04]{2004A&A...418..989N}, \citet[][VF05]{2005ApJS..159..141V}
 and \citet[][TA05]{2005PASJ...57..109T} to analyse the full samples.
 To ensure that we did not introduce any bias resulting from estimates based on
 different analysis techniques, a comparison between our metallicities and the ones reported
 in these papers is shown in Figure~\ref{valcomp}.
 Our sample contains 72 stars in common with NO04. Our metallicities are slightly higher, by
 a factor $\sim$ 0.07 dex (in median), than those given by NO04; the differences are largest
 for stars with positive metallicities. The agreement with VF05 is very good, with no apparent
 bias for the 64 stars in common; the mean difference is only -0.01 dex with a standard deviation
 of 0.09 dex. The VF05 metallicities are also higher than the NO04 values by a factor
 $\sim$ 0.08 dex. The agreement with TA05 is excellent, better than $\pm$0.10 dex for most of
 the 49 common stars. The latter result is expected because the
 same method and lines were used to estimate the metallicity; 
 it can thus be considered a consistency double check.

 \subsection{Full sample}
 \label{full_sample}

\begin{figure}
\centering
\includegraphics[angle=270,scale=0.50]{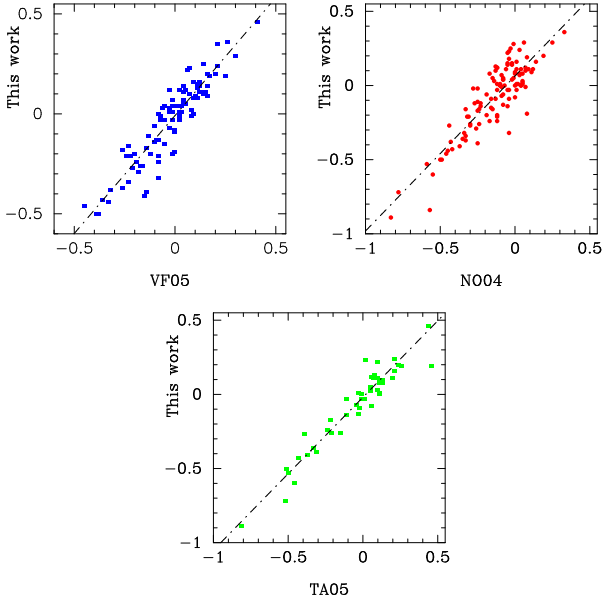}
\caption{Comparison between the metallicities from the literature and those
obtained in this work.
Top left panel: VF05;
top right panel: NO04;
bottom panel: TA05.
Dashed lines represent the best linear fit (y = m*x + b) between our metallicities and
those given in the corresponding works.
The coefficients are: 
m$=$1.18 $\pm$ 0.05, b $=$ -0.008 $\pm$ 0.008 for VF05;
m$=$1.04 $\pm$ 0.05, b $=$  0.064 $\pm$ 0.013 for N04; and
m$=$0.99 $\pm$ 0.05, b $=$ -0.017 $\pm$ 0.011 for TA05.}
\label{valcomp}
\end{figure}

 To  set  the  VF05  and  NO04  metallicities  on  our  own
 metallicity scale,  we used  the stars in  common to obtain  a linear
 transformation  (Figure~\ref{valcomp}).  Where  possible, VF05  values
 were  selected  because they  have been  obtained  from high-resolution
 spectra similar  to those  used in this  work.  The  metallicities in
 NO04 are  based on Str\"{o}mgren $uvby\beta$  photometry. The adopted
 final metallicity values for each  star of the SWD and SWOD samples
 are given in Table~\ref{sample_stars}.


 Some statistical diagnostics for the SWD and SWOD full samples are
 summarised    in    Table~\ref{caracola}. 
 Both samples have similar distributions.
 The {\it full} SWD distribution has a median of -0.04 dex, very close
 to the value obtained in the {\it homogeneous} analysis (-0.02 dex).
 In the case of the SWOD sample, the {\it full} sample has a median of -0.07 dex
 that, when compared with the value of -0.01 dex for the {\it homogeneous}
 subsample, means a difference of 0.06 dex.
 We note that 0.06 dex is of the order of the individual
 uncertainties in metallicity.  The  SWD and SWOD distributions have
 a    smaller    dispersion    when    we consider   the    whole    sample
 (Figure~\ref{histogram},  right  panel).  Using  a  K-S analysis,  we
 tested the possibility of both distributions to  differing within a
 98\% confidence  level; our results cannot exclude  that both samples
 come  from the  same  parent distribution  at  this confidence  level.
 Nevertheless,  the  likelihood  that   both  samples  are  drawn from the  same parent distribution
 diminishes significantly with respect  to the homogeneous sample case
 (9\%,  see Appendix~\ref{apendice}).  An  interesting aspect  is that
 there  seems  to  be  a  ``deficit''  of  stars  with  discs  in  the
 metallicity  range -0.50  $<$ [Fe/H]  $<$ -0.20.  This deficit  is not
 explicit  in the homogeneous  case (Figure~\ref{histogram},  see also
 Section~\ref{section_discussion} and Figure~\ref{histogram_acumuladas}).

\begin{table}
\centering
\caption{[Fe/H] statistics of the stellar samples}
\label{caracola}
\begin{scriptsize}
\begin{tabular}{lcccccc}
\hline\noalign{\smallskip}
$Sample$  &  $Mean$ & $ Median$ & $Deviation$  &  $Min$&  $Max$ & $N$ \\
\hline\noalign{\smallskip}
 SWODs   &  -0.09 & -0.07  & 0.22 & -1.12 & 0.36 & 145 \\
 SWDs    &  -0.08 & -0.04  & 0.26 & -1.49 & 0.39 & 107 \\
 SWDPs   &   0.08 &  0.05  & 0.17 & -0.34 & 0.36 &  29 \\
 SWPs    &   0.10 &  0.15  & 0.22 & -0.70 & 0.43 & 120 \\
\noalign{\smallskip}\hline\noalign{\smallskip}
\end{tabular}
\end{scriptsize}
\end{table}

 \subsection{Stars with known debris discs and planets}
 \label{secion_debris_planets}


\begin{table}
\centering
\caption{Stars with known debris discs and planets.}
\label{planetas_discos}
\begin{tabular}{lllrll}
\hline\noalign{\smallskip}
HIP     & HD            & SpType        &   [Fe/H]$^{\dag}$     & Ref$^{\ddag}$ &  Planet$^{\star}$ \\  
        &               &               &   (dex)               &               &                   \\ 
\hline\noalign{\smallskip}
522	&	142	&	F7V	&	0.09	(b)	&       9:  & gc   \\
1499	&	1461	&	G0V	&	0.18	(b)	&	11  & mlh  \\
7978	&	10647	&	 F8V	&	-0.09	(a)	&	3   & gc   \\
14954	&	19994	&	F8V	&	0.19	(a)	&	8   & gc   \\
15510   &       20794   &       G8V     &       -0.34   (a)     &       10  & mlc  \\
16537	&	22049	&	K2V	&	-0.08	(a)	&	1   & gc   \\
27253	&	38529	&	G4V	&	0.31	(b)	&	6   & mgc  \\
27435   &       38858   &       G4V     &       -0.27   (a)     &       4   & lc   \\
28767	&	40979	&	F8	&	0.13	(b)	&	10: & gc   \\
30503   &       45184   &       G2V     &       0.03    (b)     &       11  & lh   \\
31246	&	46375	&	K1IV	&	0.23	(b)	&	10: & gh   \\
32970	&	50499	&	G1V	&	0.29	(b)	&	10: & gc   \\
33212	&	50554	&	F8	&	-0.09	(b)	&	8   & gc   \\
33719$^{\sharp}$	&	52265	&	G0V     &	0.18	(b)	&	8   & gc   \\
40693	&	69830	&	K0V	&	0.00	(a)	&	2   & mlh  \\
42282	&	73526	&	G6V	&	0.22	(b)	&	10: & mgc  \\
47007	&	82943	&	G0	&	0.23	(b)	&	8   & mgc  \\
58451	&	104067	&	K2V	&	0.04	(b)	&	11  & gc   \\
61028	&	108874	&	G5	&	0.17	(b)	&	12  & mgc  \\
64924	&	115617	&	G5V	&	0.00	(a)	&	8   & mlh  \\
65721	&	117176	&	G5V	&	-0.03	(a)	&	8   & gc   \\
71395	&	128311	&	K0	&	0.04	(a)	&	8   & mgc  \\
72339	&	130322	&	K0V	&	-0.07	(b)	&	12  & gh   \\
94075   &       178911B &       G5      &       0.29    (b)     &       10: & gc   \\
97546	&	187085	&	G0V	&	0.05	(b)	&	10: & gc   \\
99711	&	192263	&	K2	&	-0.01	(a)	&	5   & gc   \\
104903	&	202206	&	G6V	&	0.36	(b)	&	10  & mgc  \\
112190  &       215152  &       K0      &       -0.10   (c)     &      11   & mlh  \\
113044	&	216435	&	G3IV	&	0.24	(b)	&	10  & gc   \\
\hline\noalign{\smallskip}
\multicolumn{6}{l}{$^{\dag}$ (a) This work; (b)  \cite{2005ApJS..159..141V}}\\
\multicolumn{6}{l}{(b) values are set into our metallicity scale as described in}\\
\multicolumn{6}{l}{Section~\ref{previous_work}.}\\
\multicolumn{6}{l}{(c) Metallicity for this star is from \cite{2008A&A...487..373S} since no}\\
\multicolumn{6}{l}{value were found in VF05 or NO04.}\\
\hline
\multicolumn{6}{l}{$^{\ddag}$ (1) \cite{Habing01}; (2) \cite{Bryden06};}\\
\multicolumn{6}{l}{(3) \cite{Moor06};  (4) \cite{Beichman06};}\\
\multicolumn{6}{l}{(5) \cite{2006ApJ...644L.125S}; (6) \cite{2007ApJ...658.1312M};}\\
\multicolumn{6}{l}{(7) \cite{Rhee07}; (8) \cite{Trilling08}; }\\
\multicolumn{6}{l}{(9) \cite{2009ApJ...705.1226B}; (10) \cite{2009ApJ...700L..73K}; }\\
\multicolumn{6}{l}{(11) \cite{2010ApJ...710L..26K}; (12) \cite{DodsonRobinson11}} \\
\multicolumn{6}{l}{The symbol ``:'' means that non-excess is attributed to the}\\
\multicolumn{6}{l}{corresponding star in (9) or (10).}\\
\hline
\multicolumn{6}{l}{$^{\star}$ m = multiplanet system; l = low-mass planet;}\\
\multicolumn{6}{l}{g = gas giant planet; c = cool planet;}\\
\multicolumn{6}{l}{h = hot planet (semimajor axis $\le$ 0.1 AU, see text);}\\
\hline
\multicolumn{6}{l}{$^{\sharp}$ Spectral type from \cite{2001MNRAS.328...45M}}\\
\noalign{\smallskip}\hline\noalign{\smallskip}
\end{tabular}
\end{table}

  At the time of writting\footnote{December  26, 2011}, there are, to our
 knowledge, 29 solar-type stars known to host both a debris disc
 and at least one planet (SWDPs). This figure represents an increase of
 $\sim$  50\% with respect  to the  most recent  works
 \citep{2009ApJ...705.1226B,2009ApJ...700L..73K}
 \footnote{For the 22 stars with debris
   discs and planets given  by \cite{2009ApJ...700L..73K}, HD 33636 has
   a substellar companion that has  been retracted as a planet (Bean
   et al. 2007), GJ 581 is a M star, and HD 137759 is a giant star. In
   addition, \cite{2009ApJ...705.1226B}  listed HD 150706 as hosting a planet
   and   a   debris   disc,   but   the  planet   is   not   confirmed
   (http://exoplanet.eu).}.     These    stars    are     listed    in
   Table~\ref{planetas_discos}.

   Among  the 29  SWDPs, 11  stars host  known multiplanet  systems, which
   represents  an incidence rate  of 38\%.  \citet[][]{2009ApJ...693.1084W}  found a
   rate of  14\% confirmed multiple planetary systems,  and it  could be
   28\%  or higher  when they  include  cases  with a  significant
   evidence    of    being   multiple\footnote{See  also   for    comparison
   http//exoplanet.eu}.   \citet{2011arXiv1109.2497M}   found   a   rate
   exceeding 70\% among  their 24 systems with planets  less massive than
   30 M$_\oplus$. In our SWDP sample,
   there are five stars with low-mass planets in multiple systems,
   but  this might  be a  lower  limit, as
   pointed  out by  \citet{2011arXiv1109.2497M}. This  suggests  that the
   multiplanet system  rate in SWDPs  approaches that of  the low-mass
   planet case.

   There are 7 of 29 SWDPs that host at  least one low-mass planet,
   M $\lesssim$ 30 M$_\oplus$.   These stars are HD 1461, HD 20794,
   HD 38858, HD 45184, HD 69830, 61 Vir (HD 115617), and HD 215152; in all cases, but
   in HD  1461, their metallicity  is [Fe/H] $\leq$ 0.0,  consistent with
   the metallicity trend for stars with low-mass planets
   \citep[e.g.][]{2011arXiv1109.2497M,2011A&A...533A.141S}.

   \citet[][Figure 9]{2009ApJ...693.1084W} and \citet[][Figure 1]{2009ApJ...694L.171C}
   showed  that there is  an enhanced frequency
   of close-in  gas giant planets with semimajor axes  $\lesssim$ 0.07 AU
   (hot Jupiters). Among  the 22 SWDPs that are  currently known to host
   only gas-giant  planets, HD  46375 is the  only star harbouring  such a
   close-in planet, semimajor axis of 0.041 AU, while HD 130322 has a hot
   Jupiter at 0.088 AU;
   five more stars have giant planets with semimajor
   axes smaller than 0.5 AU (HD  38529, HD 104067, HD 117176, HD 178911B,
   and HD 192263). On the other  hand, the semimajor axes of the low-mass
   planets are  $\lesssim$ 0.07 AU in all  cases, but in HD  20794 and HD
   38858 .

\begin{figure*}[!htb]
\centering
\includegraphics[angle=270,scale=0.45]{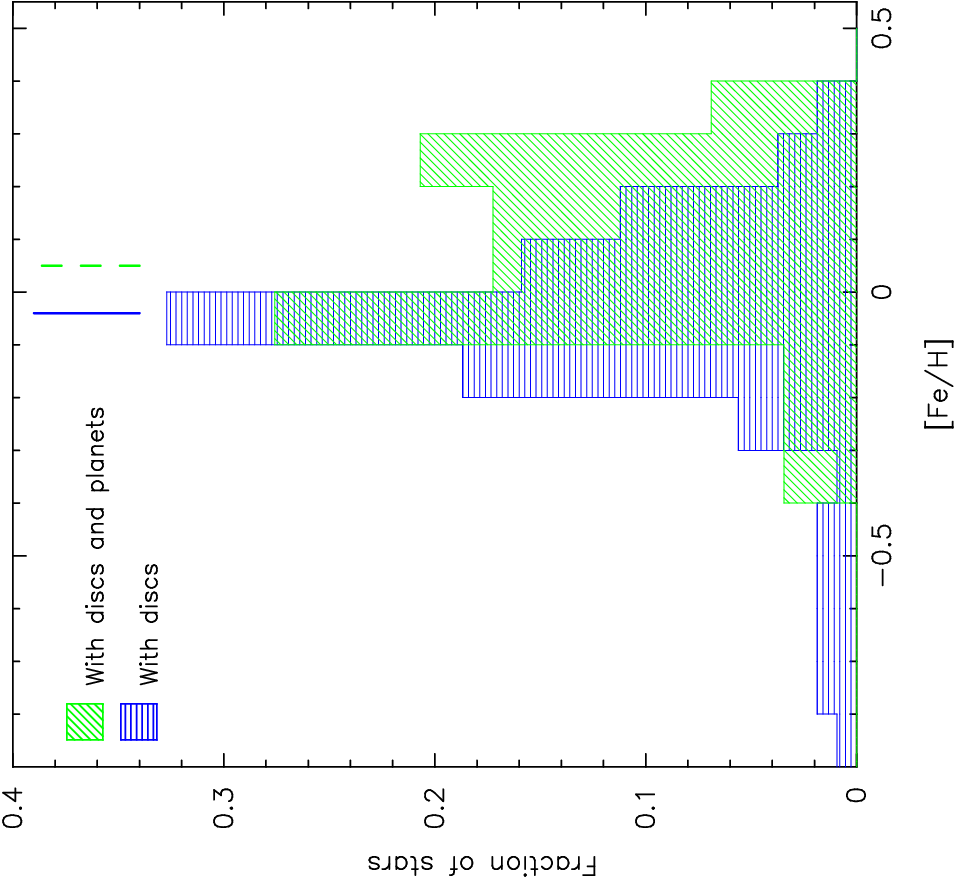}
\includegraphics[angle=270,scale=0.45]{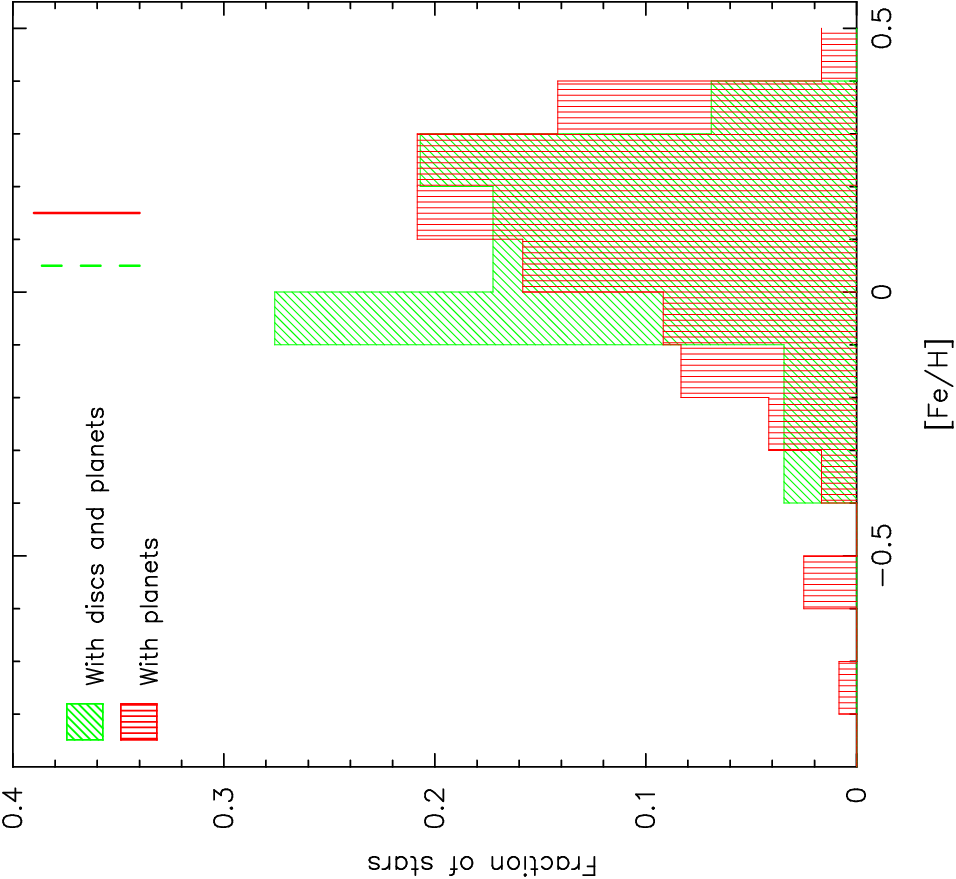}
\caption{
Normalized  metallicity  distribution  of  the  SWDP  sample (light green histogram)
versus stars  with debris discs (left) and stars with giant planets (right). Median values of the distributions are
shown with  vertical lines.
}
\label{debris_planets}
\end{figure*}

   The statistical properties of the SWDP  metallicity distribution
   are  shown  in
   Table~\ref{caracola},  while Figure~\ref{debris_planets}  (left) compares
   the  corresponding  SWDP  histogram  with the  SWDs.   The  figure
   clearly shows the distinct  metallicity distributions of both the SWDPs
   and the SWDs; a K-S  test confirms that both distributions differ within
   a  confidence  level  of  98\%  (the  likelihood  of  being  the  same
   distribution is 0.7\%).

   Summarizing, although there may be  some bias related to the planet
   detection methods as well as
   the sensitivity in detecting  debris discs, 
   our results suggest that  SWDPs
   i) have higher metallicities than both SWDs and SWODs
   (see Figures~\ref{debris_planets} and ~\ref{histogram_acumuladas}),
   ii) they  tend to have  a higher incidence
   of multiplanet systems,  most  likely at a rate close to
   the one  of stars with low-mass  planets, iii) many of them host low-mass
   planets,
   and iv) in the cases  with only gas-giant planets, these planets tend to
   be cool Jupiters 
   (only two out of 22 stars harbour one hot Jupiter).
 

\begin{figure*}[!htb]
\centering
\includegraphics[angle=270,scale=0.45]{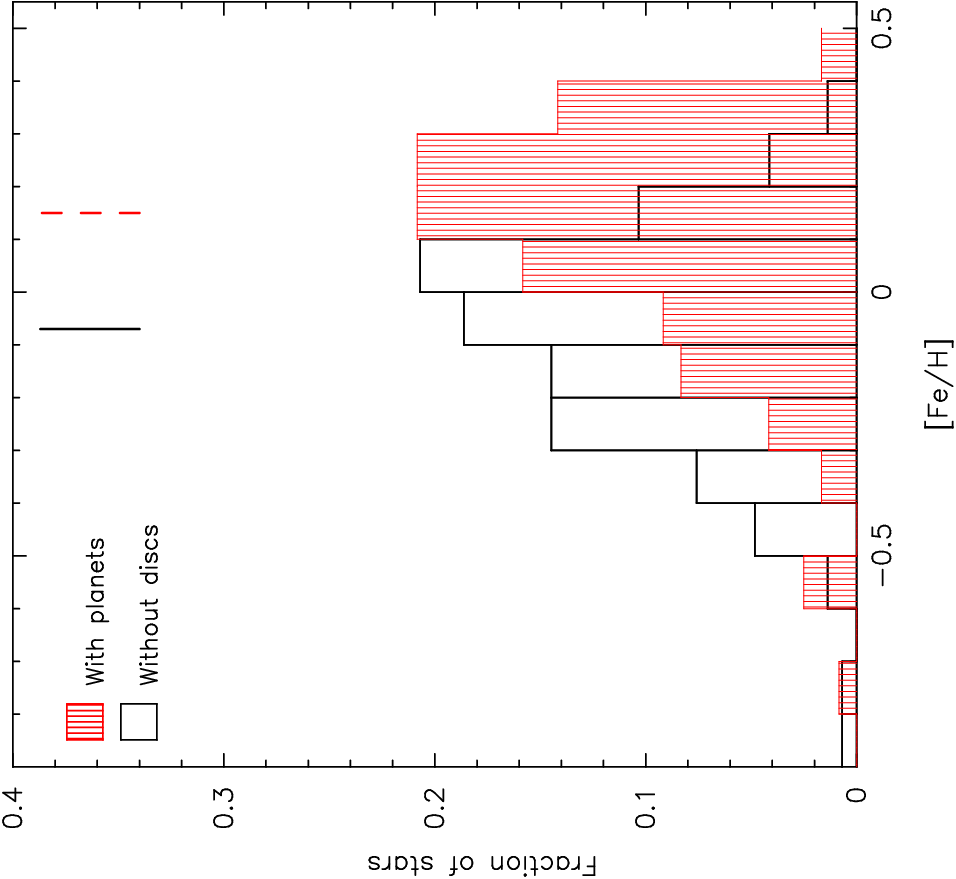}
\includegraphics[angle=270,scale=0.45]{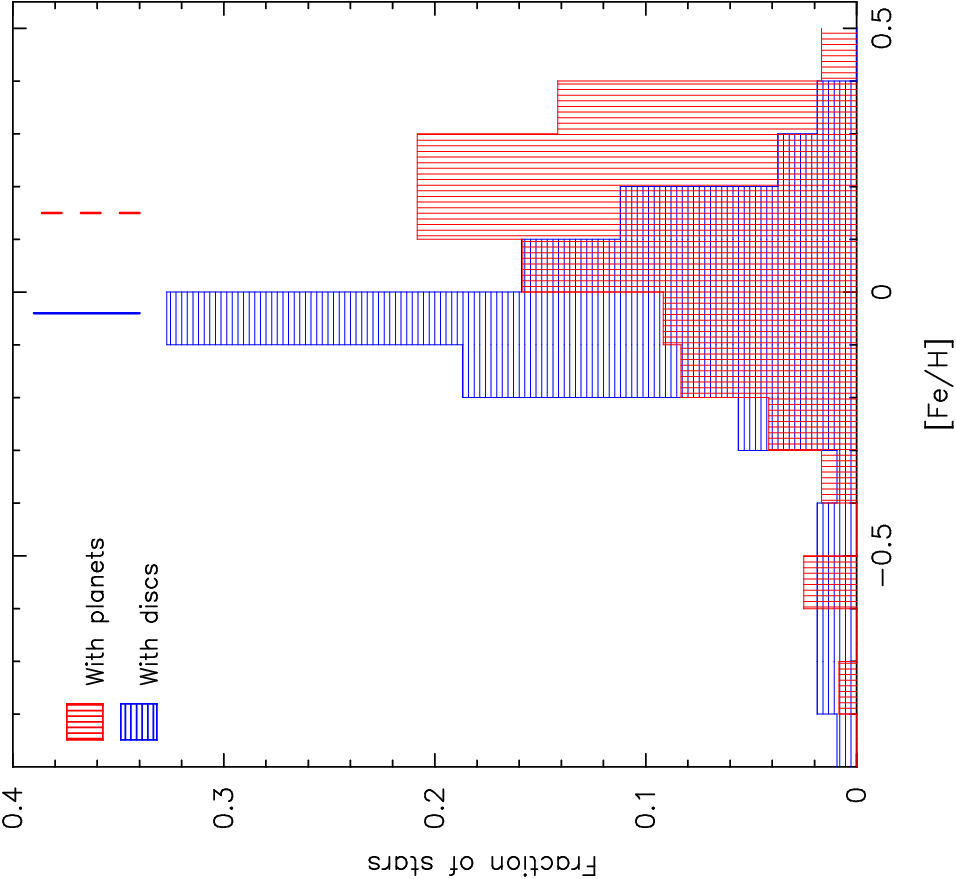}
\caption{
  Normalized metallicity distributions of planet host-stars (red
  histogram) versus  stars without debris  discs (left)  and  stars
  with debris discs (right). Median values of the distributions
  are shown with vertical lines.
}
\label{histogram_planets}
\end{figure*}

\subsection{Comparison with stars with giant planets}

 Figure~\ref{debris_planets} (right)  shows    the    metallicity
 distributions of  both SWDPs and  SWPs.  The SWPs sample  contains 120 
 stars and  corresponds to stars hosting exclusively giant  planets
 from \cite{2004A&A...415.1153S}, \cite{2005ApJS..159..141V}, \cite{2011A&A...533A.141S},
 and \cite{2011arXiv1109.2497M},
 where we  have removed
 the stars   with retracted  or not-confirmed exoplanets. 
 Both histograms clearly show that the
 stars in the SWP and SWDP samples tend to have
 high metallicity. The 
 K-S tests cannot rule out that both distributions are the same
 (p-value = 49\%).

 With the aim of completeness, Figure~\ref{histogram_planets} compares the
 metallicity distribution of the SWPs, with those of the SWODs and SWDs
 samples, where the well-known trend of
 SWPs (gas-giant planets) to higher metallicities is clearly reproduced.

\section{Discussion}
\label{section_discussion}


\begin{figure}
\centering
\includegraphics[angle=270,scale=0.45]{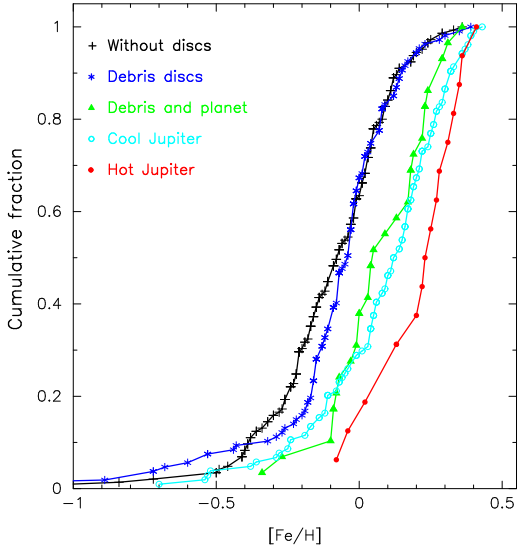}
\caption{
Histogram of cumulative frequencies for the different samples studied
in this work.}

\label{histogram_acumuladas}
\end{figure}

 The results  presented in the  previous section suggest that  a transition
 toward higher metallicities occurs from SWODs to SWPs.
 The  cumulative  metallicity
 distributions, presented in Figure~\ref{histogram_acumuladas}, allow  us to get an
 unified overview  of the metallicity trends.  As  pointed out before,
 the distribution of SWDs is similar to that of SWODs, but there seems
 to be a deficit of SWDs at low [Fe/H], below approximately -0.1 (see also
 the histogram for the  full samples in Figure~\ref{histogram} (right
 panel)  and the  median  [Fe/H] value  in Table~\ref{caracola}).   The
 distribution of SWDPs  can be clearly distinguished from that  of SWDs and is
 similar  to  that  of SWPs.  Thus, planets are clearly   the  main drivers of
 the trend in stellar  metallicity in SWDPs;
 this is true for both the low-mass and the giant planets in the SWDP
 sample.  The  metallicity distribution of SWPs was divided into
 hot and cool Jupiters  because most of the
 SWDPs hosting  giant planets are  associated with cool  planets.
 Figure ~\ref{histogram_acumuladas} suggests that the frequency of hot
 giant planets is  lower for low  metallicities than the frequency  of cool
 ones. 
 We point out
 that a similar  trend is obtained, when the  data refer to all
 known  solar-type  stars  hosting  giant planets,  i.e.,  stars  with
 close-in giant planets tend to be more metal-rich.

 These   trends   can   be   explained   by   core-accretion   models
 \citep[e.g.][]{1996Icar..124...62P,2004ApJ...616..567I,2005Icar..179..415H,
 2009A&A...501.1139M,2012arXiv1201.1036M},
  and    are
 consistent with  the view that  the mass of solids  in proto-planetary
 discs is the  main factor controlling the  formation of planets and
 planetesimals \citep{2007MNRAS.378L...1G,2007ApJ...658.1312M}.
 Thus,
 the  rapid build-up  of a  core in  a metal-rich  proto-planetary disc
 would allow giant planets to form before the  dissipation of the gas,
 while the  formation of planetesimals  could proceed slowly  after the
 gas dissipation  and also  in a less  metal-rich environment.  We note
 that planetesimals  could form regardless of the giant  planet formation, and
 that the  timescale for  Earth-like planet formation  is long  and can
 proceed in a relatively metal-poor environment.


\begin{figure}
\centering
\includegraphics[angle=270,scale=0.45]{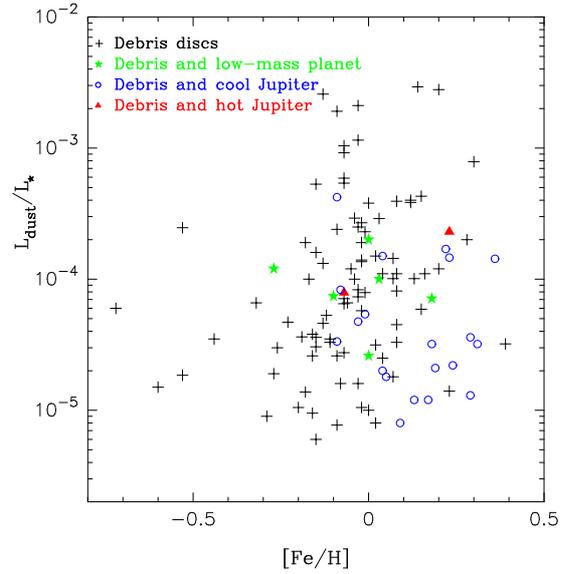}
\caption{Fractional dust luminosity, L$_{\rm dust}$/L$_{\star}$, versus [Fe/H]
for those stars hosting a debris discs. Stars are plotted with different symbols and
colours depending on the presence/absence of low-mass or cool/hot Jupiter planets.}
\label{fdust_vs_metallicity}
\end{figure}


\begin{figure}
\centering
\includegraphics[angle=270,scale=0.45]{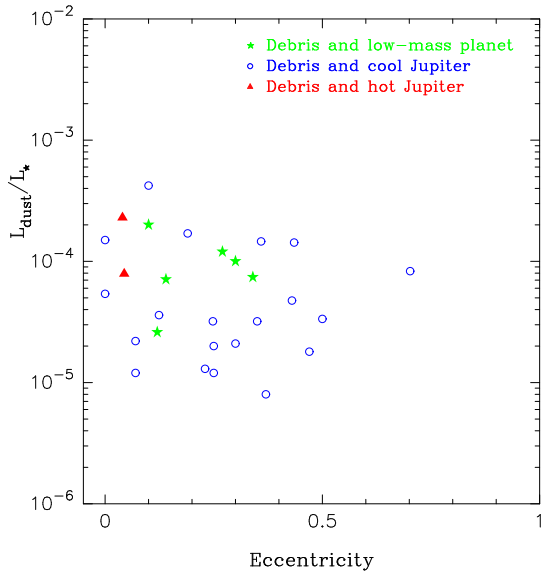}
\caption{Fractional dust luminosity, L$_{\rm dust}$/L$_{\star}$, versus
eccentricity.}
\label{fdust_vs_ecc}
\end{figure}


\begin{figure}
\centering
\includegraphics[angle=270,scale=0.45]{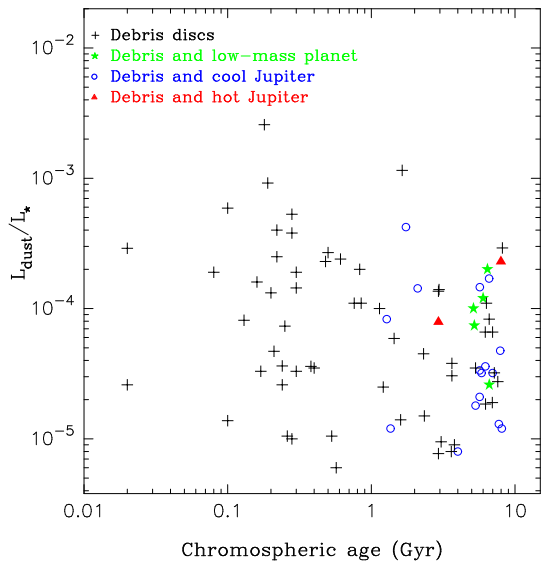}
\caption{Fractional dust luminosity, L$_{\rm dust}$/L$_{\star}$, versus stellar age.}
\label{fdust_vs_age}
\end{figure}

 Figure~\ref{fdust_vs_metallicity}    shows    the   fractional    dust
 luminosity, L$_{\rm  dust}$/L$_{\star}$, of the SWDs  and SWDPs versus
 the  metallicity.    The  plot  distinguishes   between  low-mass  and
 gas giant planets. Values of L$_{\rm dust}$/L$_{\star}$ are taken from
 the references in Section~\ref{stellar_sample}; we plot the mean value
 of L$_{\rm dust}$/L$_{\star}$ for the stars from \cite{Trilling08}. It
 is found  that the  SWDPs as a  whole span approximately two orders of
 magnitude in L$_{\rm dust}$/L$_{\star}$ and are well-mixed with SWDs, while
 most of the  stars hosting debris discs and cool giant  planets tend to have
 low  dust luminosities, L$_{\rm  dust}$/L$_{\star}$ $<$  10$^{-4}$; 
 more than 50\% of SWDPs of this type are indeed  
 concentrated in the
 low-dust luminosity/high-metallicity        corner        of
 Figure~\ref{fdust_vs_metallicity}.
 In  addition, there seems to be a trend
 of larger  eccentricities (we take as reference  the innermost planet)
 while    the    luminosity    of    the   dust    decreases,  
 albeit with a large scatter (Figure~\ref{fdust_vs_ecc}). 
 Such an  anticorrelation  may be the  result  of
 dynamical  instabilities produced  by eccentric  giant  planets, which
 clear  out  the  inner  and  outer  regions  of  the  planetary  discs
 \citep{2011A&A...530A..62R}. On the other hand, there is no trend with
 the  semimajor axis  of the  planet (not shown), although  it seems  that low-mass
 planets tend to be predominantly hot but most of the giant planets are
 cool (Section~\ref{secion_debris_planets}). 
 Furthermore,  while the SWDs span the $\sim$10 Myr  - 10 Gyr
 range, the  SWDPs are  mature stars (older  than 1 Gyr), although low-mass
 planet host stars tend  to group at old ages, $>$ 5  Gyr, and the cool
 giant-planet stars span a larger range of 1 - 10 Gyr (Figure
 \ref{fdust_vs_age}). This  age behaviour reflects a bias introduced by 
 current planet-detection
 techniques.
 Young stars are usually excluded from planet-search programmes
 owing to their high-levels of chromospheric activity,
 although much effort is being applied
 to overcome this problem 
 \citep[e.g.][]{2011A&A...527A..82D,2011A&A...525A.140D}.
 Finally, we can exclude 
 a dust luminosity evolution with age in the SWDP sample,
 in line with the results of
 \cite{Trilling08} for solar-type stars surrounded by debris discs.

\section{Conclusions}

The number of debris disc stars known to host planets has increased in
the past  few years  by a factor of $\sim$ 50  \% , particularly those
associated  with low-mass  planets.   This   has  motivated us  to
revisit the  properties of  these stars and  to compare them  with stars
with planets, stars with debris discs, and stars 
with neither debris nor planets.

We have identified  a transition toward  higher metallicities from 
SWODs to SWPs. The
SWDs have a  metallicity distribution similar to those of SWODs,
although  the distribution of the first ones
might be slightly shifted  towards higher metallicities. The SWDPs 
follow  the same  metallicity trend  as SWPs,
irrespective  of whether  the planets  are  low-mass or  gas
giants; thus,  it is  the planet which  reveals the metallicity  of the
corresponding  stars.
There is a high rate of incidence of multiplanet systems in SWDPs. 
Their innermost planets are  usually cool giants,
but  the planets are  close-in when  the debris  disc stars  only host
low-mass planets.  It cannot  be excluded that this latter  result could be
biased by the planet  detection techniques.  These results support the
scenario of core accretion for  planet formation and the previous view
that the mass of solids in proto-planetary discs is the main factor 
determining the outcome of planet formation processes.

In addition, we have found that debris disc stars hosting cool giant planets
tend  to have  the  lowest dust  luminosities,  and that  there is  an
anticorrelation between  the dust luminosity and  the innermost planet
eccentricity. A  plausible explanation  of these suggested  trends is
provided by recent simulations  of dynamical instabilities produced by
eccentric  giant  planets.   These  apparent  trends  will  likely  be either
confirmed or  rejected by the various programmes  dealing with planets
and debris discs, currently being carried out with the {\it Herschel Space
Observatory},  together  with  the  expected   detection  of  further
planets, particularly low-mass planets, around the debris disc stars.

Finally, no other trend has been found relating debris disc and planet (e.g.
period or semimajor axis) properties.

\begin{acknowledgements}

  This work was supported by the Spanish Ministerio de Ciencia e Innovaci\'on (MICINN),
  Plan Nacional de Astronom\'ia y Astrof\'isica, under grant \emph{AYA2008-01727}.
  J.M. acknowledges support from the Universidad Aut\'onoma de Madrid
  (Department of Theoretical Physics). We sincerely appreciate the careful
  reading of the manuscript and the constructive comments of an anonymous
  referee. 

\end{acknowledgements}


\bibliographystyle{aa}
\bibliography{abundances_debris.bib}


\begin{appendix}
\section{Results of the Kolmogorov-Smirnov tests}
\label{apendice}

 The Kolmogorov-Smirnov test (hereafter K-S test) is widely used to study
 the significance of the difference between two data samples \citep[e.g.][]{1983MNRAS.202..615P}. 
 It is based on the maximum deviation between the empirical
 distribution functions of both samples\\
 
 \begin{equation}
 D = max|F_{1}(x) - F_{2}(x)|, 
 \end{equation} 

 \noindent where F$_{1}(x)$ and F$_{2}(x)$ are the empirical distribution functions
 of the first and second samples respectively, and are given by

 \begin{equation}
 F(x) = \frac{n(x_{i} \le x)}{N}.  
 \end{equation}

 \noindent The K-S test tests the null hypothesis
 H$_{0}$ that F$_{1}(x)$=F$_{2}(x)$, i.e., both samples come from the
 same underlying continuous distribution, which 
 is accepted if
 \begin{equation}
  max|F_{1}(x) - F_{2}(x)| < C_{\frac{\alpha}{2},n_{1},n_{2}}
 \end{equation}

 \noindent where n$_{1}$ and n$_{2}$ are the sizes of the samples, $\alpha$ is the
 confidence level, and
 $C$ the corresponding critical values of the K-S distribution.

 Through this paper, we perform several K-S tests between the different
 samples studied. Results are given in Table~\ref{ksresults}, where the  ``asymptotic p value''
 is also given. It provides an estimate
 of the likelihood of the null hypothesis and
 is reasonable accurate for samples sizes for which

 \begin{equation}
 n_{\rm eff} = \frac{n_{1} \times n_{2}}{n_{1} + n_{2}} \ge 4  .
 \end{equation} 

 \noindent All the tests were made at a confidence level $\alpha$=0.02.


\begin{table}
\centering
\caption{Results of the K-S tests performed in this work.}
\label{ksresults}
\begin{scriptsize}
\begin{tabular}{llrrrccc}
\hline\noalign{\smallskip}
Sample 1 & Sample 2 &  n$_{1}$ &  n$_{2}$ & n$_{\rm eff}$  & H$_{0}$$^{\ddag}$  & $p$ & $D$ \\
\hline\noalign{\smallskip}
SWDs$^{\star}$   & SWODs$^{\star}$   &   35 &  58 & 22   & 0  & 0.94               & 0.11  \\
SWDs$^{\sharp}$  & SWODs$^{\sharp}$  &  107 & 145 & 62   & 0  & 0.09               & 0.16  \\
\hline
SWDPs            & SWODs             &   29 & 145 & 24   & 1  & $\sim$10$^{-3}$    & 0.40  \\
SWDPs            & SWDs              &   29 & 107 & 23   & 1  & 7$\times$10$^{-3}$ & 0.34  \\
SWDPs            & SWPs              &   29 & 120 & 24   & 0  & 0.49               & 0.17  \\
\hline
SWPs             & SWODs             &  120 & 145 & 66   & 1  & $\sim$10$^{-11}$  & 0.43  \\
SWPs             & SWDs              &  120 & 107 & 57   & 1  & $\sim$10$^{-10}$  & 0.44  \\
\hline
\multicolumn{8}{l}{$\ddag$ 0: Accept null hypothesis; 1: Reject null hypothesis}\\
\multicolumn{8}{l}{$\star$: Homogeneous samples; $\sharp$: Full samples}\\
\multicolumn{8}{l}{$\dag$: Only stars with giant planets considered}\\
\noalign{\smallskip}\hline\noalign{\smallskip}
\end{tabular}
\end{scriptsize}
\end{table}

\end{appendix}


\Online
\section{Tables}
\label{tables}

  Results produced in the framework of this work 
  are only available 
  in the electronic version of the corresponding paper or
  at the CDS via anonymous ftp to cdsarc.u-strasbg.fr (130.79.128.5)
  or via {\tt http://cdsweb.u-strasbg.fr/cgi-bin/qcat?J/A+A/}

 Table~\ref{sample_stars} lists the stars in the SWDs (stars with known debris discs) and
 SWODs (stars without debris discs) samples, as well as their properties:
 HIP number (column 1); HD number (column 2); Hipparcos spectral-type (column 3);
 distance in parsec (column 4); $\log$Age(yr) (column 5); [Fe/H] and its reference
 (column 6); and references for debris disc detection (column 7). 
 References for [Fe/H] are:  
 (a) this work; (b) \cite{2005ApJS..159..141V}; (c) \cite{2004A&A...418..989N}; 
 metallicities taken from (b) and (c) are set on the metallicity scale of this work
 as described in Section~\ref{previous_work}.
 References in column 6 are as follows:
 (1) \cite{Habing01}; (2) \cite{Spangler01}; (3) \cite{2005ApJ...634.1372C};
 (4) \cite{Beichman06}; (5) \cite{Bryden06}; (6) \cite{Moor06};
 (7) \cite{2006ApJ...644L.125S}; (8) \cite{Rhee07}; (9) \cite{Trilling08};  
 (10) \cite{2009ApJ...700L..73K}; (11) \cite{0004-637X-698-2-1068}; (12) \cite{Tanner2009ApJ};
 (13) \cite{2010ApJ...710L..26K}; (14) \cite{2011ApJS..193....4M}.

 Table~\ref{parameters_table} contains:
 HIP number (column 1); HD number (column 2); Hipparcos spectral-type (column 3);
 effective temperature in kelvin (column 4); logarithm of the surface gravity in cms$^{\rm -2}$ (column 5);
 microturbulent velocity in km$s^{\rm -1}$ (column 6); final metallicity in dex (column 7);
 mean iron abundance derived from Fe~{\sc I} lines (column 8) in the usual scale
 ($ A(Fe) = \log[(N_{Fe}/N_{H}) + 12$]); number of Fe~{\sc I} lines used (column 9);
 mean iron abundance derived from Fe~{\sc II} lines (column 10);
 number of Fe~{\sc II} lines used (column 11); and spectrograph (column 12).
 Each measured quantity is accompanied by its corresponding uncertainty.


\onllongtabL{1}{
\label{sample_stars_full}
\begin{small}
\begin{longtable}{llllllrl}
\multicolumn{8}{c}{{ \tablename\ \thetable{} The SWDs and SWODs samples.}}\\
\hline
HIP & HD & SpType & V & distance & $\log$(Age) & [Fe/H]$^{\dag}$  & Ref \\
    &    &        &   & (pc)     &   (yr)      &  (dex)           &     \\
\hline
\endfirsthead
\multicolumn{8}{c}{{ \tablename\ \thetable{}  Continued}}\\
\hline
HIP & HD & SpType & V & distance & $\log$(Age) & [Fe/H]$^{\dag}$  & Ref \\
    &    &        &   & (pc)     &   (yr)      &  (dex)           &     \\
\hline
\endhead
\hline
\endfoot
\endlastfoot
\hline\noalign{\smallskip}
\multicolumn{8}{c}{Stars with known debris discs.}\\
\hline\noalign{\smallskip}
171	&	224930	&	G3V	&	5.80	&	12.17	&	9.60	&	-0.72	(a)	&	13	\\
490	&	105	&	 G0V	&	7.51	&	39.39	&	8.34	&	-0.03	(b)	&	2	\\
544	&	166	&	K0V	&	6.07	&	13.67	&	9.16	&	0.15	(a)	&	5	\\
682	&	377	&	 G2V	&	7.59	&	39.08	&	8.34	&	0.12	(b)	&	6	\\
1481	&	1466	&	F8/G0	&	7.46	&	41.55	&	8.34	&	-0.22	(c)	&	7	\\
1598	&	1562	&	G0	&	6.97	&	24.80	&	9.79	&	-0.32	(a)	&	13	\\
1599	&	1581	&	F9V	&	4.23	&	8.59	&	9.58	&	-0.29	(a)	&	9	\\
2843	&	3296	&	F5	&	6.72	&	45.05	&		&	0.02	(c)	&	9	\\
3810	&	4676	&	F8V	&	5.07	&	23.45	&	9.71	&	0.00	(c)	&	13	\\
4148	&	5133	&	K2V	&	7.15	&	14.17	&	9.56	&	-0.16	(b)	&	13	\\
5336	&	6582	&	G5V	&	5.17	&	7.55	&	9.69	&	-0.89	(a)	&	13	\\
5862	&	7570	&	F8V	&	4.97	&	15.10	&	9.62	&	0.17	(a) 	&	10	\\
5944	&	7590	&	G0	&	6.59	&	23.20	&	8.70	&	-0.02	(a)	&	11	\\
6878	&	8907	&	 F8	&	6.66	&	34.77	&	8.78	&	-0.09	(b)	&	6	\\
7576	&	10008	&	G5	&	7.66	&	23.96	&	8.10	&	0.08	(a)	&	11	\\
8102	&	10700	&	G8V	&	3.49	&	3.65	&	9.77	&	-0.43	(a)	&	1	\\
11160	&	15060	&	F5	&	7.02	&	75.99	&		&	-0.08	(c)	&	14	\\
12623	&	16739	&	F9V	&	4.91	&	24.19	&	9.83	&	0.21	(c)	&	13	\\
13402	&	17925	&	K1V	&	6.05	&	10.35	&	8.24	&	0.08	(a)	&	1	\\
13642	&	18143	&	K2	&	7.52	&	23.52	&		&	0.35	(a)	&	13	\\
15371	&	20807	&	G1V	&	5.24	&	12.03	&	9.49	&	-0.16	(a)	&	9	\\
16852	&	22484	&	F9V	&	4.29	&	13.97	&	9.88	&	-0.07	(a)	&	9	\\
17439	&	23484	&	K1V	&	6.99	&	16.03	&	8.88	&	0.04	(b)	&	13	\\
18859	&	25457	&	 F5V	&	5.38	&	18.84	&	9.06	&	-0.04	(c)	&	6	\\
19335	&	25998	&	F7V	&	5.52	&	21.00	&	9.36	&	0.08	(c)	&	4	\\
22263	&	30495	&	G3V	&	5.49	&	13.27	&	9.08	&	0.04	(a)	&	1	\\
22295	&	32195	&	F7V	&	8.14	&	61.01	&		&	-0.07	(c)	&	14	\\
23693	&	33262	&	F7V	&	4.71	&	11.64	&	8.76	&	-0.15	(c)	&	5	\\
23816	&	33081	&	F7V	&	7.04	&	50.61	&		&	-0.13	(c)	&	14	\\
24205	&	33636	&	G0	&	7.00	&	28.37	&	9.56	&	-0.15	(b)	&	9	\\
24947	&	35114	&	F6V	&	7.39	&	48.31	&		&	-0.12	(c)	&	14	\\
25486	&	35850	&	F7V	&	6.30	&	27.04	&	7.29	&	-0.09	(b)	&	2	\\
26779	&	37394	&	K1V	&	6.21	&	12.28	&	8.93	&	0.14	(a)	&	12	\\
27072	&	38393	& 	 F7V	&	3.59	&	8.93	&	9.47	&	-0.09	(a)	&	8	\\
27980	&	39833	&  G0III	&	7.65	&	41.22	&		&	0.20	(c)	&	8	\\
29568	&	43162	&	G5V	&	6.37	&	16.72	&	8.45	&	0.00	(a)	&	10	\\
31711	&	48189	&	G1/G2V	&	6.15	&	21.29	&	8.02	&	-0.18	(c)	&	11	\\
32480	&	48682	&	G0V	&	5.24	&	16.72	&	9.80	&	0.16	(a)	&	4	\\
32775	&	50571	& F7III-IV	&	6.11	&	33.62	&	8.93	&	0.08	(c)	&	6	\\
33690	&	53143	&	 K0IV-V	&	6.81	&	18.33	&	8.92	&	0.28	(a)	&	6	\\
34819	&	55052	&	  F5IV	&	5.85	&	109.77	&		&	0.07	(c)	&	8	\\
36515	&	59967	&	G3V	&	6.66	&	21.82	&	8.39	&	-0.19	(c)	&	11	\\
36827	&	60491	&	K2V	&	8.16	&	24.56	&	7.92	&	-0.18	(a)	&	13	\\
36906	&	60234	&	  G0	&	7.68	&	133.87	&		&	0.15	(c)	&	8	\\
36948	&	61005	&	  G3/G5V	&	8.23	&	35.35	&	8.27	&	-0.13	(c)	&	8	\\
42333	&	73350	&	G0	&	6.74	&	23.98	&	8.48	&	0.07	(a)	&	11	\\
42430	&	73752	&	  G3/G5V	&	5.05	&	19.38	&	9.86	&	0.39	(c)	&	8	\\
42438	&	72905	&	G1.5V	&	5.63	&	14.35	&	8.41	&	-0.02	(a)	&	2	\\
43625	&	75616	&	F5	&	6.92	&	35.42	&		&	-0.26	(c)	&	9	\\
43726	&	76151	&	G3V	&	6.01	&	17.39	&	9.20	&	0.23	(a)	&	12	\\
50384	&	89125	&	F8V	&	5.81	&	22.82	&	9.84	&	-0.27	(a)	&	13	\\
52462	&	92945	&	K1V	&	7.72	&	21.40	&	8.45	&	-0.15	(b)	&	3	\\
56830	&	101259	&	G6/G8V	&	6.40	&	63.09	&		&	-0.72	(b)	&	9	\\
59422	&	105912	&	F5	&	6.95	&	49.68	&		&	-0.01	(c)	&	9	\\
60025	&	107067	&	F8	&	8.69	&	65.96	&	9.22	&	-0.03	(c)	&	2	\\
60074	&	107146	&	 G2V	&	7.04	&	27.46	&	8.28	&	-0.07	(b)	&	6	\\
60582	&	108102	&	F8	&	8.12	&	95.15	&	8.00	&	-0.07	(c)	&	2	\\
62207	&	110897	&	G0V	&	5.95	&	17.38	&	9.80	&	-0.53	(a)	&	9	\\
63584	&	113337	&	  F6V	&	6.01	&	36.89	&		&	0.13	(c)	&	8	\\
66704	&	119124	&	F8V	&	6.31	&	25.33	&	8.38	&	-0.16	(c)	&	3	\\
66765	&	118972	&	K1V	&	6.92	&	15.65	&	8.60	&	-0.11	(b)	&	9	\\
66781	&	119332	&	K0IV-V	&	7.77	&	24.63	&		&	-0.07	(a)	&	13	\\
68101	&	121384	&	  G8V	&	6.00	&	38.70	&		&	-0.53	(b)	&	8	\\
68380	&	122106	&	 F8V	&	6.36	&	77.52	&		&	0.20	(c)	&	6	\\
68593	&	122652	&	  F8	&	7.16	&	39.29	&	9.47	&	-0.02	(b)	&	8	\\
69682	&	124718	&	  G5V	&	8.89	&	63.17	&		&	-0.03	(c)	&	8	\\
69989	&	125451	&	F5IV	&	5.41	&	26.10	&		&	0.07	(c)	&	2	\\
70344	&	126265	&	  G2III	&	7.21	&	70.42	&		&	0.12	(c)	&	8	\\
72848	&	131511	&	K2V	&	6.00	&	11.51	&	8.84	&	0.13	(a)	&	13	\\
73869	&	134319	&	G5	&	8.40	&	44.74	&	8.20	&	-0.15	(b)	&	2	\\
74702	&	135599	&	K0	&	6.92	&	15.85	&	8.31	&	-0.13	(a)	&	11	\\
74975	&	136202	&	F8III-IV	&	5.04	&	25.38	&	9.92	&	0.02	(c)	&	13	\\
76375	&	139323	&	  K3V	&	7.65	&	22.38	&		&	0.30	(b)	&	8	\\
76635	&	139590	&	  G0V	&	7.50	&	55.77	&		&	0.08	(c)	&	8	\\
76829	&	139664	&	F5IV-V	&	4.64	&	17.44	&		&	-0.05	(c)	&	1	\\
79492	&	145958	&	G8V	&	6.68	&	23.79	&	9.73	&	-0.09	(b)	&	13	\\
81800	&	151044	&	  F8V	&	6.48	&	29.33	&	9.82	&	-0.03	(b)	&	1	\\
85235	&	158633	&	K0V	&	6.44	&	12.80	&	9.73	&	-0.44	(a)	&	4	\\
88399	&	164249	&	 F5V	&	7.01	&	48.15	&		&	-0.07	(c)	&	6	\\
88745	&	165908	&	F7V	&	5.05	&	15.64	&	9.37	&	-0.60	(a)	&	13	\\
89770	&	169666	&	 F5	&	6.68	&	53.22	&		&	0.02	(c)	&	6	\\
90936	&	170773	&	 F5V	&	6.22	&	36.98	&	8.44	&	0.00	(c)	&	6	\\
93815	&	177171	&	F7	&	5.17	&	56.72	&		&	0.01	(c)	&	7	\\
94050	&	177996	&	K1V	&	7.89	&	33.84	&	8.33	&	-0.23	(c)	&	2	\\
94858	&	180134	&	F7	&	6.36	&	45.07	&		&	-0.12	(c)	&	7	\\
95270	&	181327	&	 F5/F6V	&	7.04	&	51.81	&		&	0.14	(c)	&	6	\\
96258	&	184960	&	F7V	&	5.71	&	25.11	&		&	-0.17	(c)	&	2	\\
99273	&	191089	&	 F5V	&	7.18	&	52.22	&		&	-0.09	(c)	&	6	\\
99316	&	191499	&	K0	&	7.56	&	23.64	&		&	-0.16	(a)	&	13	\\
102626	&	197890	&	K0V	&	9.44	&	52.19	&		&	-1.49	(c)	&	11	\\
103389	&	199260	&	F7V	&	5.70	&	21.97	&	8.48	&	-0.11	(c)	&	4	\\
104239	&	200968	&	K1V	&	7.12	&	17.58	&	9.47	&	-0.01	(a)	&	13	\\
105184	&	202628	&	G5V	&	6.75	&	24.42	&	9.47	&	-0.02	(b)	&	13	\\
105388	&	202917	&	 G5V	&	8.65	&	42.97	&	7.18	&	0.03	(b)	&	2	\\
107022	&	205536	&	  G8V	&	7.07	&	22.02	&	9.91	&	-0.04	(b)	&	8	\\
107350	&	206860	&	G0V	&	5.96	&	17.88	&	8.73	&	-0.20	(a)	&	9	\\
107412	&	206893	&	 F5V	&	6.69	&	38.34	&	8.68	&	-0.01	(c)	&	6	\\
107649	&	207129	&	G2V	&	5.57	&	15.99	&	9.84	&	-0.06	(b)	&	1	\\
108028	&	208038	&	K0	&	8.18	&	23.04	&	8.48	&	-0.02	(a)	&	13	\\
108809	&	209253	&	  F6/F7V	&	6.63	&	30.14	&	8.39	&	-0.03	(b)	&	2	\\
110753	&	212695	&	F5	&	6.94	&	46.49	&		&	-0.02	(c)	&	9	\\
111170	&	213429	&	F7V	&	6.15	&	25.41	&	9.56	&	0.02	(c)	&	14	\\
114236	&	218340	&	G3	&	8.44	&	56.59	&		&	0.09	(a)	&	7	\\
114924	&	219623	&	F7V	&	5.58	&	20.50	&		&	-0.03	(c)	&	4	\\
114948	&	219482	&	F7V	&	5.64	&	20.54	&	8.59	&	-0.15	(c)	&	4	\\
117712	&	223778	&	K3V	&	6.36	&	10.89	&	8.93	&	-0.68	(c)	&	13	\\
	&	3670	&	F5V	&	8.23	&		&		&	-0.07	(c)	&	14	\\
\hline\noalign{\smallskip}
\multicolumn{8}{c}{Stars without debris discs.}\\
\hline\noalign{\smallskip}
394	&	225239	&	G2V	&	6.09	&	39.18	&		&	-0.41	(c)	&		\\
462	&	63	&	F5	&	7.13	&	50.66	&		&	-0.09	(c)	&		\\
910	&	693	&	F5V	&	4.89	&	18.75	&	9.48	&	-0.32	(c)	&		\\
1573	&	1539	&	F5	&	7.03	&	43.69	&		&	-0.04	(c)	&		\\
2802	&	3302	&	F6V	&	5.51	&	34.82	&	8.50	&	0.04	(c)	&		\\
3170	&	3823	&	G1V	&	5.89	&	24.96	&	9.63	&	-0.26	(b)	&		\\
3185	&	3795	&	G3/G5V	&	6.14	&	28.89	&	9.82	&	-0.49	(b)	&		\\
3236	&	3861	&	F5	&	6.52	&	33.44	&	9.25	&	0.05	(b)	&		\\
3559	&	4307	&	G2V	&	6.15	&	31.00	&	9.89	&	-0.22	(b)	&		\\
3765	&	4628	&	K2V	&	5.74	&	7.45	&		&	-0.24	(a)	&		\\
3909	&	4813	&	F7IV-V	&	5.17	&	15.75	&	2.93	&	-0.16	(a)	&		\\
7601	&	10800	&	G2V	&	5.88	&	27.38	&	9.09	&	-0.11	(c)	&		\\
7981	&	10476	&	K1V	&	5.24	&	7.53	&		&	-0.03	(a)	&		\\
8486	&	11131	&	G0	&	6.72	&	22.56	&	7.72	&	-0.03	(a)	&		\\
10306	&	13555	&	F5V	&	5.23	&	28.87	&		&	-0.21	(c)	&		\\
10798	&	14412	&	G8V	&	6.33	&	12.67	&	9.61	&	-0.46	(a)	&		\\
11072	&	14802	&	G2V	&	5.19	&	21.96	&	9.80	&	-0.06	(a)	&		\\
11548	&	15335	&	G0V	&	5.89	&	31.41	&		&	-0.24	(b)	&		\\
11783	&	15798	&	F5V	&	4.74	&	26.70	&	9.26	&	-0.26	(c)	&		\\
12114	&	16160	&	K3V	&	5.79	&	7.18	&		&	-0.19	(a)	&		\\
12777	&	16895	&	F7V	&	4.10	&	11.13	&	9.90	&	0.12	(a)	&		\\
14632	&	19373	&	G0V	&	4.05	&	10.54	&	9.82	&	0.11	(a)	&		\\
15330	&	20766	&	G2V	&	5.53	&	12.01	&	9.61	&	-0.21	(a)	&		\\
15457	&	20630	&	G5Vvar	&	4.84	&	9.14	&	8.54	&	0.09	(a)	&		\\
17378	&	23249	&	K0IV	&	3.52	&	9.04	&		&	0.03	(b)	&		\\
19855	&	26913	&	G5IV	&	6.94	&	21.06	&	8.45	&	-0.11	(c)	&		\\
22449	&	30652	&	F6V	&	3.19	&	8.07	&	9.48	&	-0.01	(b)	&		\\
23311	&	32147	&	K3V	&	6.22	&	8.71	&		&	0.29	(a)	&		\\
24786	&	34721	&	G0V	&	5.96	&	25.03	&	9.79	&	-0.24	(a)	&		\\
24813	&	34411	&	G0V	&	4.69	&	12.63	&	9.83	&	0.08	(a)	&		\\
25278	&	35296	&	F8V	&	5.00	&	14.39	&	8.41	&	-0.09	(c)	&		\\
27913	&	39587	&	G0V	&	4.39	&	8.66	&	8.75	&	-0.09	(a)	&		\\
28954	&	41593	&	K0	&	6.76	&	15.27	&	8.70	&	0.19	(a)	&		\\
29271	&	43834	&	G5V	&	5.08	&	10.20	&	9.74	&	0.12	(a)	&		\\
32984	&	50281	&	K3V	&	6.58	&	8.71	&		&	-0.02	(b)	&		\\
33277	&	50692	&	G0V	&	5.74	&	17.24	&	9.74	&	-0.11	(a)	&		\\
34017	&	52711	&	G4V	&	5.93	&	19.13	&	9.74	&	-0.14	(a)	&		\\
35136	&	55575	&	G0V	&	5.54	&	16.89	&	9.72	&	-0.36	(a)	&		\\
36439	&	58855	&	F6V	&	5.35	&	20.24	&		&	-0.23	(c)	&		\\
37283	&	60912	&	F5	&	6.89	&	45.50	&		&	-0.07	(c)	&		\\
37853	&	63077	&	G0V	&	5.36	&	15.21	&	9.55	&	-0.72	(c)	&		\\
38172	&	63333	&	F5	&	7.09	&	43.08	&		&	-0.38	(c)	&		\\
39780	&	67228	&	G2IV	&	5.30	&	23.29	&	9.69	&	0.19	(b)	&		\\
39903	&	68456	&	F5V	&	4.74	&	19.98	&	8.09	&	-0.17	(c)	&		\\
40843	&	69897	&	F6V	&	5.13	&	18.27	&		&	-0.39	(a)	&		\\
41226	&	70843	&	F5	&	7.06	&	46.30	&		&	0.14	(b)	&		\\
41484	&	71148	&	G5V	&	6.32	&	22.25	&		&	0.01	(a)	&		\\
41573	&	71640	&	F5	&	7.40	&	44.74	&		&	-0.18	(c)	&		\\
41926	&	72673	&	K0V	&	6.38	&	12.21	&	9.75	&	-0.40	(b)	&		\\
42074	&	72760	&	G5	&	7.32	&	21.14	&	8.03	&	0.01	(a)	&		\\
42808	&	74576	&	K2V	&	6.58	&	11.14	&	8.73	&	-0.16	(a)	&		\\
44728	&	77967	&	F0	&	6.61	&	43.25	&		&	-0.36	(c)	&		\\
44897	&	78366	&	F9V	&	5.95	&	19.19	&	8.70	&	0.03	(b)	&		\\
45699	&	80218	&	F5	&	6.61	&	40.65	&		&	-0.21	(c)	&		\\
47403	&	83451	&	F5	&	7.12	&	49.04	&		&	-0.06	(c)	&		\\
47436	&	83525	&	F5	&	6.90	&	48.97	&		&	-0.07	(c)	&		\\
47592	&	84117	&	G0V	&	4.93	&	15.01	&	9.62	&	-0.21	(a)	&		\\
48113	&	84737	&	G2V	&	5.08	&	18.37	&	9.87	&	0.10	(a)	&		\\
48768	&	86147	&	F5	&	6.70	&	47.01	&		&	-0.01	(c)	&		\\
50366	&	88984	&	F5	&	7.30	&	51.15	&		&	-0.21	(c)	&		\\
51459	&	90839	&	F8V	&	4.82	&	12.78	&	9.47	&	-0.15	(a)	&		\\
52574	&	93081	&	F5	&	7.09	&	48.38	&		&	-0.22	(c)	&		\\
53252	&	94388	&	F6V	&	5.23	&	30.81	&	9.54	&	0.24	(c)	&		\\
54745	&	97334	&	G0V	&	6.41	&	21.93	&	8.03	&	-0.01	(a)	&		\\
55666	&	99126	&	F5	&	6.94	&	49.75	&		&	-0.09	(c)	&		\\
56186	&	100067	&	F5	&	7.17	&	39.95	&		&	-0.32	(c)	&		\\
56452	&	100623	&	K0V	&	5.96	&	9.56	&	9.61	&	-0.38	(a)	&		\\
56997	&	101501	&	G8Vvar	&	5.31	&	9.61	&	8.91	&	0.01	(a)	&		\\
57507	&	102438	&	G5V	&	6.48	&	17.47	&	9.72	&	-0.28	(b)	&		\\
57757	&	102870	&	F8V	&	3.59	&	10.93	&	9.81	&	0.09	(a)	&		\\
57939	&	103095	&	G8Vp	&	6.42	&	9.09	&	9.67	&	-1.12	(a)	&		\\
58268	&	103773	&	F5	&	6.73	&	46.45	&		&	0.05	(c)	&		\\
58803	&	104731	&	F6V	&	5.15	&	25.32	&	8.56	&	-0.01	(c)	&		\\
61100	&	109011	&	K2V	&	8.08	&	25.10	&		&	-0.34	(a)	&		\\
61578	&	109756	&	F5	&	6.95	&	48.52	&		&	-0.15	(c)	&		\\
62523	&	111395	&	G7V	&	6.29	&	16.93	&	8.57	&	0.22	(a)	&		\\
62636	&	111545	&	F5	&	6.99	&	47.76	&		&	-0.02	(c)	&		\\
63033	&	112164	&	G2IV	&	5.89	&	40.10	&		&	0.20	(c)	&		\\
63742	&	113449	&	G5V	&	7.69	&	21.70	&	8.28	&	-0.17	(a)	&		\\
64394	&	114710	&	G0V	&	4.23	&	9.13	&	9.89	&	0.11	(a)	&		\\
64408	&	114613	&	G3V	&	4.85	&	20.67	&		&	0.18	(b)	&		\\
64792	&	115383	&	G0Vs	&	5.19	&	17.56	&	8.61	&	0.24	(a)	&		\\
65515	&	116956	&	G9IV-V	&	7.29	&	21.59	&	7.84	&	0.03	(a)	&		\\
65530	&	117043	&	G6V	&	6.50	&	21.17	&		&	0.29	(a)	&		\\
67195	&	120005	&	F5	&	6.51	&	43.73	&		&	0.05	(c)	&		\\
67620	&	120690	&	G5V	&	6.43	&	19.47	&	9.31	&	0.05	(a)	&		\\
69040	&	123691	&	F2	&	6.80	&	53.02	&		&	-0.08	(c)	&		\\
69090	&	122862	&	G1V	&	6.02	&	28.50	&	9.82	&	-0.15	(b)	&		\\
70497	&	126660	&	F7V	&	4.04	&	14.53	&	9.03	&	-0.08	(c)	&		\\
70873	&	127334	&	G5V	&	6.36	&	23.74	&		&	0.25	(b)	&		\\
71743	&	128987	&	G6V	&	7.24	&	23.67	&	8.81	&	0.01	(a)	&		\\
72130	&	130460	&	F5	&	7.22	&	49.12	&		&	-0.01	(c)	&		\\
72567	&	130948	&	G2V	&	5.86	&	18.17	&	8.28	&	-0.01	(b)	&		\\
72573	&	133002	&	F9V	&	5.63	&	43.29	&		&	-0.41	(c)	&		\\
73996	&	134083	&	F5V	&	4.93	&	19.55	&	9.60	&	0.05	(b)	&		\\
74605	&	136064	&	F9IV	&	5.15	&	25.34	&	9.46	&	-0.03	(c)	&		\\
77372	&	141128	&	F5	&	7.00	&	49.24	&		&	-0.19	(c)	&		\\
77408	&	141272	&	G8V	&	7.44	&	21.29	&	8.29	&	-0.14	(a)	&		\\
77760	&	142373	&	F9V	&	4.60	&	15.89	&	9.87	&	-0.50	(a)	&		\\
77801	&	142267	&	G0IV	&	6.07	&	17.35	&	9.60	&	-0.50	(a)	&		\\
77838	&	143105	&	F5	&	6.76	&	48.66	&		&	0.03	(c)	&		\\
78072	&	142860	&	F6V	&	3.85	&	11.25	&		&	-0.17	(b)	&		\\
79672	&	146233	&	G1V	&	5.49	&	13.90	&	9.63	&	0.04	(a)	&		\\
81300	&	149661	&	K2V	&	5.77	&	9.75	&	9.36	&	0.05	(a)	&		\\
82588	&	152391	&	G8V	&	6.65	&	17.25	&	8.44	&	0.04	(a)	&		\\
84862	&	157214	&	G0V	&	5.38	&	14.33	&	9.77	&	-0.41	(a)	&		\\
88972	&	166620	&	K2V	&	6.38	&	11.02	&	9.76	&	-0.07	(b)	&		\\
89348	&	168151	&	F5V	&	4.99	&	22.92	&	9.62	&	-0.17	(c)	&		\\
90485	&	169830	&	F8V	&	5.90	&	36.60	&	9.81	&	0.09	(b)	&		\\
91120	&	171886	&	F5	&	7.16	&	49.43	&		&	-0.30	(c)	&		\\
91438	&	172051	&	G5V	&	5.85	&	13.08	&	9.66	&	-0.21	(a)	&		\\
92043	&	173667	&	F6V	&	4.19	&	19.21	&	9.68	&	0.12	(b)	&		\\
93252	&	176441	&	F5	&	7.06	&	46.77	&		&	-0.16	(c)	&		\\
93858	&	177565	&	G8V	&	6.15	&	16.95	&	9.78	&	0.07	(b)	&		\\
94346	&	180161	&	G8V	&	7.04	&	20.02	&	8.84	&	0.18	(a)	&		\\
94981	&	181655	&	G8V	&	6.29	&	25.39	&	9.79	&	0.02	(b)	&		\\
95149	&	181321	&	G1/G2V	&	6.48	&	18.83	&	8.39	&	-0.21	(c)	&		\\
96100	&	185144	&	K0V	&	4.67	&	5.75	&	9.56	&	-0.26	(a)	&		\\
96895	&	186408	&	G2V	&	5.99	&	21.08	&	9.84	&	0.02	(a)	&		\\
97675	&	187691	&	F8V	&	5.12	&	19.19	&	9.85	&	0.13	(b)	&		\\
98066	&	188376	&	G3/G5III	&	4.70	&	25.99	&	9.87	&	0.06	(c)	&		\\
98819	&	190406	&	G1V	&	5.80	&	17.77	&	9.50	&	0.12	(a)	&		\\
98959	&	189567	&	G2V	&	6.07	&	17.73	&	9.61	&	-0.22	(b)	&		\\
99240	&	190248	&	G5IV-Vvar	&	3.55	&	6.11	&	9.82	&	0.36	(a)	&		\\
99461	&	191408	&	K2V	&	5.32	&	6.02	&		&	-0.40	(b)	&		\\
100017	&	193664	&	G3V	&	5.91	&	17.57	&	9.58	&	-0.14	(b)	&		\\
101983	&	196378	&	F8V	&	5.11	&	24.66	&	9.57	&	-0.39	(b)	&		\\
101997	&	196761	&	G8/K0V	&	6.36	&	14.38	&	9.72	&	-0.30	(b)	&		\\
102485	&	197692	&	F5V	&	4.13	&	14.68	&	8.49	&	0.03	(c)	&		\\
103931	&	200433	&	F5	&	6.91	&	47.37	&		&	-0.05	(c)	&		\\
105202	&	202884	&	F5	&	7.27	&	41.81	&		&	-0.27	(c)	&		\\
105858	&	203608	&	F6V	&	4.21	&	9.26	&	8.76	&	-0.84	(a)	&		\\
109422	&	210302	&	F6V	&	4.94	&	18.28	&	9.55	&	0.09	(b)	&		\\
109821	&	210918	&	G5V	&	6.23	&	22.05	&		&	-0.09	(b)	&		\\
110649	&	212330	&	F9V	&	5.31	&	20.56	&		&	0.00	(b)	&		\\
112447	&	215648	&	F7V	&	4.20	&	16.30	&		&	-0.20	(b)	&		\\
113829	&	217813	&	G5V	&	6.65	&	24.72	&		&	0.02	(b)	&		\\
114622	&	219134	&	K3Vvar	&	5.57	&	6.55	&		&	0.10	(b)	&		\\
115220	&	219983	&	F2	&	6.64	&	48.73	&		&	-0.12	(c)	&		\\
115331	&	220182	&	K1V	&	7.36	&	21.52	&	7.56	&	0.11	(a)	&		\\
116250	&	221420	&	G2V	&	5.82	&	31.44	&		&	0.33	(b)	&		\\
116613	&	222143	&	G5	&	6.58	&	23.33	&	9.08	&	0.14	(a)	&		\\
116745	&	222237	&	K3V	&	7.09	&	11.42	&		&	-0.24	(b)	&		\\
116771	&	222368	&	F7V	&	4.13	&	13.71	&		&	-0.13	(a)	&		\\
116906	&	222582	&	G5	&	7.68	&	41.77	&		&	-0.03	(b)	&		\\
\noalign{\smallskip}\hline\noalign{\smallskip}
\end{longtable}
\end{small}
}


\onllongtabL{5}{
\begin{center}
\label{metallicities_table_full}
\begin{tiny}
\begin{longtable}{lllccccccccc}
\multicolumn{12}{c}{{ \tablename\ \thetable{}  Basic physical parameters and metallicities for the stars measured in this work.}}\\
\hline
HIP &  HD &  SpType & T$_{\rm eff}$ &  $\log g$      & $\xi_{t}$     & [Fe/H] &  $\left\langle A({\rm Fe}~{\rm I}) \right\rangle$ & n$_{\rm I}$ &
 $\left\langle A({\rm Fe}~{\rm II}) \right\rangle$  & n$_{\rm II}$ &  Spec.$^{\dag}$ \\
    &     &     & (K)   & (${\rm cm s^{-2}}$)  & (${\rm km s^{-1}}$) & dex  &      &      &      &      &      \\
 (1)& (2) & (3) & (4)   & (5)            & (6)           & (7)  & (8)  & (9)  & (10) & (11) & (12) \\
\hline
\endfirsthead
\multicolumn{12}{c}{{ \tablename\ \thetable{}  Continued}}\\
\hline
HIP &  HD &  SpType & T$_{\rm eff}$ &  $\log g$      & $\xi_{t}$     & [Fe/H] &  $\left\langle A({\rm Fe}~{\rm I}) \right\rangle$ & n$_{\rm I}$ &
 $\left\langle A({\rm Fe}~{\rm II}) \right\rangle$  & n$_{\rm II}$ &  Spec.$^{\dag}$ \\
    &     &     & (K)   & (${\rm cm s^{-2}}$)  & (${\rm km s^{-1}}$) & dex  &      &      &      &      &      \\
 (1)& (2) & (3) & (4)   & (5)            & (6)           & (7)  & (8)  & (9)  & (10) & (11) & (12) \\
\hline
\endhead
\hline
\endfoot
\endlastfoot
\multicolumn{12}{c}{\textbf{Stars with known debris discs}}\\
\noalign{\smallskip}\hline\noalign{\smallskip}
171	&	224930	&	G3V	&	5491	$\pm$	31	&	4.75	$\pm$	0.12	&	0.92	$\pm$	0.40	&	-0.72	$\pm$	0.08	&	6.78	$\pm$	0.11	&	52	&	6.78	$\pm$	0.12	&	12	&	4	\\
544	&	166	&	K0V	&	5575	$\pm$	51	&	4.68	$\pm$	0.14	&	1.05	$\pm$	0.25	&	0.15	$\pm$	0.05	&	7.65	$\pm$	0.06	&	57	&	7.65	$\pm$	0.06	&	13	&	4	\\
1598	&	1562	&	G0	&	5603	$\pm$	36	&	4.30	$\pm$	0.12	&	0.67	$\pm$	0.27	&	-0.32	$\pm$	0.06	&	7.18	$\pm$	0.07	&	58	&	7.18	$\pm$	0.09	&	12	&	1	\\
1599	&	1581	&	F9V	&	5809	$\pm$	39	&	4.24	$\pm$	0.12	&	1.30	$\pm$	0.30	&	-0.29	$\pm$	0.06	&	7.21	$\pm$	0.08	&	59	&	7.22	$\pm$	0.10	&	13	&	5	\\
5336	&	6582	&	G5V	&	5291	$\pm$	32	&	4.57	$\pm$	0.11	&	0.82	$\pm$	0.42	&	-0.89	$\pm$	0.08	&	6.61	$\pm$	0.11	&	55	&	6.62	$\pm$	0.12	&	12	&	4	\\
5862	&	7570	&	F8V	&	6111	$\pm$	35	&	4.42	$\pm$	0.10	&	1.35	$\pm$	0.17	&	0.17	$\pm$	0.03	&	7.67	$\pm$	0.04	&	50	&	7.67	$\pm$	0.04	&	13	&	6	\\
5944	&	7590	&	G0	&	5951	$\pm$	39	&	4.65	$\pm$	0.11	&	1.04	$\pm$	0.38	&	-0.02	$\pm$	0.05	&	7.48	$\pm$	0.06	&	48	&	7.48	$\pm$	0.08	&	11	&	1	\\
7576	&	10008	&	G5	&	5293	$\pm$	68	&	4.90	$\pm$	0.19	&	0.39	$\pm$	0.45	&	0.08	$\pm$	0.06	&	7.58	$\pm$	0.07	&	53	&	7.58	$\pm$	0.10	&	9	&	1	\\
8102	&	10700	&	G8V	&	5312	$\pm$	137	&	4.59	$\pm$	0.13	&	0.15	$\pm$	0.69	&	-0.43	$\pm$	0.15	&	7.07	$\pm$	0.24	&	63	&	7.07	$\pm$	0.20	&	11	&	4	\\
13402	&	17925	&	K1V	&	5103	$\pm$	47	&	4.51	$\pm$	0.17	&	0.87	$\pm$	0.22	&	0.08	$\pm$	0.06	&	7.58	$\pm$	0.05	&	45	&	7.58	$\pm$	0.09	&	12	&	4	\\
13642	&	18143	&	K2V	&	5162	$\pm$	54	&	4.54	$\pm$	0.15	&	0.31	$\pm$	0.38	&	0.35	$\pm$	0.03	&	7.86	$\pm$	0.03	&	53	&	7.85	$\pm$	0.05	&	10	&	1	\\
15371	&	20807	&	G1V	&	5874	$\pm$	40	&	4.64	$\pm$	0.11	&	0.87	$\pm$	0.27	&	-0.16	$\pm$	0.05	&	7.34	$\pm$	0.06	&	61	&	7.34	$\pm$	0.07	&	13	&	5	\\
16852	&	22484	&	F9V	&	5979	$\pm$	56	&	4.68	$\pm$	0.17	&	1.22	$\pm$	0.19	&	-0.07	$\pm$	0.03	&	7.43	$\pm$	0.03	&	59	&	7.43	$\pm$	0.04	&	13	&	4	\\
22263	&	30495	&	G3V	&	5852	$\pm$	25	&	4.64	$\pm$	0.08	&	0.94	$\pm$	0.22	&	0.04	$\pm$	0.03	&	7.54	$\pm$	0.04	&	63	&	7.54	$\pm$	0.04	&	13	&	4	\\
26779	&	37394	&	K1V	&	5265	$\pm$	44	&	4.69	$\pm$	0.12	&	0.56	$\pm$	0.38	&	0.14	$\pm$	0.05	&	7.64	$\pm$	0.06	&	56	&	7.63	$\pm$	0.08	&	11	&	4	\\
27072	&	38393	&	F7V	&	6259	$\pm$	36	&	4.44	$\pm$	0.09	&	1.51	$\pm$	0.16	&	-0.09	$\pm$	0.04	&	7.41	$\pm$	0.05	&	33	&	7.41	$\pm$	0.05	&	10	&	5	\\
29568	&	43162	&	G5V	&	5619	$\pm$	40	&	4.65	$\pm$	0.12	&	1.04	$\pm$	0.23	&	0.00	$\pm$	0.04	&	7.50	$\pm$	0.06	&	56	&	7.50	$\pm$	0.06	&	12	&	6	\\
32480	&	48682	&	G0V	&	6132	$\pm$	26	&	4.67	$\pm$	0.07	&	1.10	$\pm$	0.18	&	0.16	$\pm$	0.03	&	7.66	$\pm$	0.03	&	55	&	7.66	$\pm$	0.04	&	11	&	1	\\
33690	&	53143	&	K0IV-V	&	5521	$\pm$	43	&	4.65	$\pm$	0.11	&	0.73	$\pm$	0.26	&	0.28	$\pm$	0.04	&	7.78	$\pm$	0.05	&	57	&	7.79	$\pm$	0.05	&	11	&	6	\\
36827	&	60491	&	K2V	&	5079	$\pm$	61	&	4.59	$\pm$	0.15	&	0.59	$\pm$	0.43	&	-0.18	$\pm$	0.10	&	7.32	$\pm$	0.12	&	45	&	7.31	$\pm$	0.17	&	11	&	1	\\
42333	&	73350	&	G0	&	5818	$\pm$	90	&	4.42	$\pm$	0.27	&	1.34	$\pm$	0.30	&	0.07	$\pm$	0.07	&	7.57	$\pm$	0.10	&	56	&	7.57	$\pm$	0.11	&	12	&	1	\\
42438	&	72905	&	G1.5Vb	&	5893	$\pm$	53	&	4.49	$\pm$	0.18	&	1.30	$\pm$	0.27	&	-0.02	$\pm$	0.05	&	7.48	$\pm$	0.07	&	50	&	7.48	$\pm$	0.08	&	13	&	1	\\
43726	&	76151	&	G3V	&	5859	$\pm$	24	&	4.77	$\pm$	0.07	&	0.65	$\pm$	0.30	&	0.23	$\pm$	0.03	&	7.73	$\pm$	0.04	&	57	&	7.73	$\pm$	0.03	&	12	&	1	\\
50384	&	89125	&	F8V	&	6118	$\pm$	49	&	4.29	$\pm$	0.14	&	1.15	$\pm$	0.23	&	-0.27	$\pm$	0.06	&	7.23	$\pm$	0.08	&	29	&	7.23	$\pm$	0.09	&	11	&	1	\\
62207	&	110897	&	G0V	&	5789	$\pm$	55	&	4.29	$\pm$	0.17	&	1.29	$\pm$	0.44	&	-0.53	$\pm$	0.10	&	6.97	$\pm$	0.13	&	53	&	6.97	$\pm$	0.15	&	11	&	1	\\
66781	&	119332	&	K0IV-V	&	5154	$\pm$	90	&	4.55	$\pm$	0.21	&	0.34	$\pm$	0.44	&	-0.07	$\pm$	0.11	&	7.44	$\pm$	0.12	&	51	&	7.43	$\pm$	0.18	&	8	&	1	\\
72848	&	131511	&	K2V	&	5319	$\pm$	50	&	4.73	$\pm$	0.13	&	0.57	$\pm$	0.38	&	0.13	$\pm$	0.05	&	7.63	$\pm$	0.06	&	53	&	7.63	$\pm$	0.07	&	13	&	4	\\
74702	&	135599	&	K0	&	5277	$\pm$	70	&	4.17	$\pm$	0.23	&	1.15	$\pm$	0.19	&	-0.13	$\pm$	0.10	&	7.37	$\pm$	0.08	&	45	&	7.37	$\pm$	0.17	&	4	&	2	\\
85235	&	158633	&	K0V	&	5210	$\pm$	44	&	4.51	$\pm$	0.13	&	0.02	$\pm$	0.49	&	-0.44	$\pm$	0.07	&	7.06	$\pm$	0.07	&	63	&	7.06	$\pm$	0.11	&	11	&	4	\\
88745	&	165908	&	F7V	&	5938	$\pm$	38	&	4.17	$\pm$	0.10	&	1.51	$\pm$	0.27	&	-0.60	$\pm$	0.07	&	6.90	$\pm$	0.09	&	41	&	6.90	$\pm$	0.11	&	11	&	3	\\
99316	&	191499	&	G8V	&	5220	$\pm$	62	&	4.42	$\pm$	0.15	&	0.51	$\pm$	0.37	&	-0.16	$\pm$	0.09	&	7.35	$\pm$	0.12	&	57	&	7.34	$\pm$	0.13	&	11	&	1	\\
104239	&	200968	&	K1V	&	5239	$\pm$	115	&	4.67	$\pm$	0.28	&	1.10	$\pm$	0.51	&	-0.01	$\pm$	0.13	&	7.49	$\pm$	0.17	&	55	&	7.49	$\pm$	0.21	&	10	&	1	\\
107350	&	206860	&	G0V	&	5750	$\pm$	36	&	4.27	$\pm$	0.12	&	1.47	$\pm$	0.32	&	-0.20	$\pm$	0.06	&	7.30	$\pm$	0.08	&	30	&	7.30	$\pm$	0.08	&	4	&	1	\\
108028	&	208038	&	K0	&	4965	$\pm$	27	&	4.58	$\pm$	0.09	&	0.61	$\pm$	0.26	&	-0.02	$\pm$	0.02	&	7.48	$\pm$	0.03	&	59	&	7.47	$\pm$	0.05	&	10	&	1	\\
114236	&	218340	&	G3V	&	5888	$\pm$	30	&	4.48	$\pm$	0.08	&	0.89	$\pm$	0.18	&	0.09	$\pm$	0.03	&	7.59	$\pm$	0.04	&	49	&	7.59	$\pm$	0.04	&	13	&	6	\\
\noalign{\smallskip}\hline\noalign{\smallskip}
\multicolumn{12}{c}{\textbf{Stars with known planets and debris discs}}\\
\noalign{\smallskip}\hline\noalign{\smallskip}
7978	&	10647	&	F8V	&	6101	$\pm$	40	&	4.49	$\pm$	0.11	&	1.38	$\pm$	0.16	&	-0.09	$\pm$	0.04	&	7.41	$\pm$	0.05	&	32	&	7.41	$\pm$	0.06	&	9	&	6	\\
14954	&	19994	&	F8V	&	6140	$\pm$	31	&	4.35	$\pm$	0.09	&	1.44	$\pm$	0.12	&	0.19	$\pm$	0.03	&	7.69	$\pm$	0.03	&	34	&	7.69	$\pm$	0.04	&	8	&	3	\\
15510   &       20794   &       G8V     &       5386   $\pm$   31    &       4.53    $\pm$   0.08    &       0.26    $\pm$   0.40    &       -0.34   $\pm$   0.06    &       7.16    $\pm$   0.09    &       56      &       7.16    $\pm$   0.08    &       12      &       5       \\
16537	&	22049	&	K2V	&	5061	$\pm$	25	&	4.65	$\pm$	0.09	&	0.37	$\pm$	0.35	&	-0.08	$\pm$	0.05	&	7.42	$\pm$	0.05	&	55	&	7.41	$\pm$	0.08	&	12	&	4	\\
27435   &       38858   &       G4V     &       5660   $\pm$   20    &       4.36    $\pm$   0.06    &       0.97    $\pm$   0.11    &       -0.27   $\pm$   0.03    &       7.23    $\pm$   0.03    &       42      &       7.23    $\pm$   0.04    &       6       &       2       \\
40693	&	69830	&	K0V	&	5400	$\pm$	40	&	4.57	$\pm$	0.10	&	0.48	$\pm$	0.44	&	0.00	$\pm$	0.05	&	7.50	$\pm$	0.07	&	63	&	7.50	$\pm$	0.07	&	13	&	4	\\
64924	&	115617	&	G5V	&	5400	$\pm$	40	&	4.57	$\pm$	0.10	&	0.48	$\pm$	0.44	&	0.00	$\pm$	0.05	&	7.50	$\pm$	0.07	&	63	&	7.50	$\pm$	0.07	&	13	&	4	\\
65721	&	117176	&	G5V	&	5546	$\pm$	39	&	4.09	$\pm$	0.12	&	1.13	$\pm$	0.21	&	-0.03	$\pm$	0.05	&	7.47	$\pm$	0.07	&	59	&	7.47	$\pm$	0.07	&	11	&	1	\\
71395	&	128311	&	K0	&	4906	$\pm$	9	&	4.32	$\pm$	0.03	&	0.88	$\pm$	0.06	&	0.04	$\pm$	0.02	&	7.54	$\pm$	0.02	&	61	&	7.54	$\pm$	0.02	&	12	&	1	\\
99711	&	192263	&	K0V	&	4920	$\pm$	49	&	4.65	$\pm$	0.13	&	0.66	$\pm$	0.30	&	-0.01	$\pm$	0.07	&	7.50	$\pm$	0.06	&	61	&	7.48	$\pm$	0.13	&	10	&	1	\\
\noalign{\smallskip}\hline\noalign{\smallskip}
\multicolumn{12}{c}{\textbf{Stars with known planets but without known debris discs}}\\
\noalign{\smallskip}\hline\noalign{\smallskip}
3093	&	3651	&	K0V	&	5196	$\pm$	31	&	4.49	$\pm$	0.11	&	0.01	$\pm$	0.35	&	0.20	$\pm$	0.03	&	7.71	$\pm$	0.03	&	53	&	7.70	$\pm$	0.05	&	13	&	4	\\
6379	&	7924	&	K0	&	5233	$\pm$	43	&	4.50	$\pm$	0.16	&	0.54	$\pm$	0.21	&	-0.20	$\pm$	0.07	&	7.30	$\pm$	0.04	&	54	&	7.30	$\pm$	0.13	&	9	&	3	\\
7513	&	9826	&	F8V	&	6153	$\pm$	42	&	4.28	$\pm$	0.12	&	1.47	$\pm$	0.20	&	0.11	$\pm$	0.04	&	7.61	$\pm$	0.05	&	49	&	7.61	$\pm$	0.06	&	13	&	4	\\
10138	&	13445	&	K0V	&	5168	$\pm$	55	&	4.56	$\pm$	0.12	&	0.49	$\pm$	0.44	&	-0.20	$\pm$	0.05	&	7.32	$\pm$	0.06	&	57	&	7.32	$\pm$	0.08	&	12	&	5	\\
49699	&	87883	&	K0V	&	5000	$\pm$	30	&	4.62	$\pm$	0.07	&	0.30	$\pm$	0.22	&	0.14	$\pm$	0.03	&	7.65	$\pm$	0.03	&	62	&	7.62	$\pm$	0.05	&	12	&	1	\\
43587   &       75732   &       G8V     &       6140    $\pm$   31      &       4.35    $\pm$   0.09    &       1.44    $\pm$   0.12    &       0.19    $\pm$   0.03    &       7.69    $\pm$   0.03    &       42      &       7.69    $\pm$   0.04    &       10      &       4       \\
53721	&	95128	&	G0V	&	5789	$\pm$	33	&	4.26	$\pm$	0.11	&	1.07	$\pm$	0.23	&	-0.03	$\pm$	0.04	&	7.47	$\pm$	0.06	&	61	&	7.47	$\pm$	0.06	&	13	&	4	\\
57443	&	102365	&	G3/G5V	&	5524	$\pm$	40	&	4.29	$\pm$	0.13	&	0.81	$\pm$	0.30	&	-0.37	$\pm$	0.07	&	7.13	$\pm$	0.08	&	56	&	7.13	$\pm$	0.09	&	12	&	5	\\
78459	&	143761	&	G2V	&	5793	$\pm$	25	&	4.28	$\pm$	0.08	&	0.93	$\pm$	0.17	&	-0.17	$\pm$	0.05	&	7.34	$\pm$	0.05	&	46	&	7.34	$\pm$	0.06	&	12	&	1	\\
79248	&	145675	&	K0V	&	5312	$\pm$	72	&	4.43	$\pm$	0.24	&	0.52	$\pm$	0.50	&	0.46	$\pm$	0.06	&	7.97	$\pm$	0.11	&	54	&	7.96	$\pm$	0.10	&	11	&	1	\\
80337	&	147513	&	G3/G5V	&	5917	$\pm$	35	&	4.61	$\pm$	0.11	&	0.96	$\pm$	0.28	&	0.10	$\pm$	0.04	&	7.60	$\pm$	0.05	&	64	&	7.60	$\pm$	0.05	&	13	&	5	\\
83389	&	154345	&	G8V	&	5461	$\pm$	35	&	4.59	$\pm$	0.09	&	0.57	$\pm$	0.46	&	-0.06	$\pm$	0.05	&	7.44	$\pm$	0.07	&	53	&	7.44	$\pm$	0.07	&	8	&	3	\\
95319	&	182488	&	G8V	&	5452	$\pm$	63	&	4.64	$\pm$	0.16	&	0.66	$\pm$	0.37	&	0.16	$\pm$	0.06	&	7.66	$\pm$	0.07	&	56	&	7.66	$\pm$	0.08	&	10	&	3	\\
96901	&	186427	&	G5V	&	5680	$\pm$	62	&	4.37	$\pm$	0.21	&	0.89	$\pm$	0.16	&	0.01	$\pm$	0.06	&	7.51	$\pm$	0.06	&	53	&	7.51	$\pm$	0.09	&	10	&	3	\\
109378	&	210277	&	G0	&	5531	$\pm$	30	&	4.25	$\pm$	0.09	&	0.83	$\pm$	0.14	&	0.20	$\pm$	0.03	&	7.70	$\pm$	0.04	&	43	&	7.70	$\pm$	0.04	&	4	&	2	\\
113357	&	217014	&	G5V	&	5710	$\pm$	20	&	4.15	$\pm$	0.16	&	1.10	$\pm$	0.09	&	0.11	$\pm$	0.03	&	7.61	$\pm$	0.03	&	52	&	7.61	$\pm$	0.06	&	10	&	3	\\
\noalign{\smallskip}\hline\noalign{\smallskip}
\multicolumn{12}{c}{\textbf{Comparison sample}}\\
\noalign{\smallskip}\hline\noalign{\smallskip}
3765	&	4628	&	K2V	&	5014	$\pm$	37	&	4.67	$\pm$	0.11	&	0.35	$\pm$	0.40	&	-0.24	$\pm$	0.02	&	7.27	$\pm$	0.04	&	61	&	7.26	$\pm$	0.07	&	12	&	4	\\
3909	&	4813	&	F7IV-V	&	6150	$\pm$	29	&	4.27	$\pm$	0.08	&	1.33	$\pm$	0.17	&	-0.16	$\pm$	0.04	&	7.34	$\pm$	0.04	&	43	&	7.34	$\pm$	0.06	&	9	&	3	\\
7981	&	10476	&	K1V	&	5262	$\pm$	69	&	4.65	$\pm$	0.17	&	0.71	$\pm$	0.50	&	-0.03	$\pm$	0.09	&	7.47	$\pm$	0.12	&	55	&	7.48	$\pm$	0.13	&	12	&	4	\\
8486	&	11131	&	G0V	&	5864	$\pm$	29	&	4.63	$\pm$	0.07	&	0.94	$\pm$	0.17	&	-0.03	$\pm$	0.03	&	7.47	$\pm$	0.04	&	57	&	7.48	$\pm$	0.04	&	13	&	6	\\
10798	&	14412	&	G8V	&	5359	$\pm$	25	&	4.59	$\pm$	0.07	&	0.48	$\pm$	0.39	&	-0.46	$\pm$	0.05	&	7.04	$\pm$	0.06	&	62	&	7.04	$\pm$	0.06	&	11	&	4	\\
11072	&	14802	&	G2V	&	5853	$\pm$	49	&	3.99	$\pm$	0.15	&	1.28	$\pm$	0.15	&	-0.06	$\pm$	0.05	&	7.44	$\pm$	0.06	&	43	&	7.44	$\pm$	0.08	&	9	&	6	\\
12114	&	16160	&	K3V	&	4857	$\pm$	52	&	4.54	$\pm$	0.12	&	0.04	$\pm$	0.43	&	-0.19	$\pm$	0.10	&	7.32	$\pm$	0.09	&	59	&	7.30	$\pm$	0.16	&	9	&	1	\\
12777	&	16895	&	F7V	&	6304	$\pm$	76	&	4.54	$\pm$	0.20	&	1.38	$\pm$	0.35	&	0.12	$\pm$	0.05	&	7.62	$\pm$	0.06	&	43	&	7.62	$\pm$	0.07	&	12	&	4	\\
14632	&	19373	&	G0V	&	5975	$\pm$	67	&	4.14	$\pm$	0.21	&	1.32	$\pm$	0.16	&	0.11	$\pm$	0.05	&	7.61	$\pm$	0.06	&	60	&	7.61	$\pm$	0.08	&	13	&	4	\\
15330	&	20766	&	G2V	&	5719	$\pm$	29	&	4.63	$\pm$	0.08	&	0.87	$\pm$	0.18	&	-0.21	$\pm$	0.04	&	7.29	$\pm$	0.05	&	54	&	7.29	$\pm$	0.06	&	13	&	5	\\
15457	&	20630	&	G5Vvar	&	5718	$\pm$	50	&	4.52	$\pm$	0.12	&	0.94	$\pm$	0.26	&	0.09	$\pm$	0.04	&	7.59	$\pm$	0.06	&	61	&	7.59	$\pm$	0.06	&	13	&	4	\\
23311	&	32147	&	K3V	&	4746	$\pm$	30	&	4.55	$\pm$	0.09	&	0.50	$\pm$	0.25	&	0.29	$\pm$	0.02	&	7.80	$\pm$	0.02	&	60	&	7.79	$\pm$	0.05	&	10	&	4	\\
24786	&	34721	&	G0V	&	5716	$\pm$	70	&	3.54	$\pm$	0.25	&	1.22	$\pm$	0.27	&	-0.24	$\pm$	0.09	&	7.26	$\pm$	0.12	&	39	&	7.26	$\pm$	0.14	&	5	&	2	\\
24813	&	34411	&	G0V	&	5871	$\pm$	36	&	4.40	$\pm$	0.11	&	1.24	$\pm$	0.19	&	0.08	$\pm$	0.04	&	7.58	$\pm$	0.05	&	62	&	7.59	$\pm$	0.05	&	13	&	4	\\
27913	&	39587	&	G0V	&	5813	$\pm$	26	&	4.45	$\pm$	0.08	&	1.12	$\pm$	0.18	&	-0.09	$\pm$	0.04	&	7.41	$\pm$	0.05	&	43	&	7.41	$\pm$	0.05	&	12	&	4	\\
28954	&	41593	&	K0	&	5515	$\pm$	31	&	4.67	$\pm$	0.09	&	1.01	$\pm$	0.32	&	0.19	$\pm$	0.04	&	7.69	$\pm$	0.06	&	52	&	7.70	$\pm$	0.05	&	12	&	1	\\
29271	&	43834	&	G8V	&	5649	$\pm$	37	&	4.60	$\pm$	0.10	&	0.80	$\pm$	0.31	&	0.12	$\pm$	0.05	&	7.62	$\pm$	0.06	&	62	&	7.62	$\pm$	0.06	&	13	&	5	\\
33277	&	50692	&	G0V	&	5870	$\pm$	31	&	4.57	$\pm$	0.09	&	0.83	$\pm$	0.23	&	-0.11	$\pm$	0.04	&	7.39	$\pm$	0.05	&	54	&	7.39	$\pm$	0.05	&	11	&	1	\\
34017	&	52711	&	G4V	&	5797	$\pm$	37	&	4.20	$\pm$	0.11	&	0.98	$\pm$	0.18	&	-0.14	$\pm$	0.05	&	7.36	$\pm$	0.06	&	62	&	7.36	$\pm$	0.07	&	13	&	1	\\
35136	&	55575	&	G0V	&	5821	$\pm$	46	&	4.29	$\pm$	0.13	&	1.32	$\pm$	0.22	&	-0.36	$\pm$	0.06	&	7.14	$\pm$	0.07	&	59	&	7.14	$\pm$	0.09	&	13	&	1	\\
40843	&	69897	&	F6V	&	6041	$\pm$	69	&	3.79	$\pm$	0.18	&	1.51	$\pm$	0.35	&	-0.39	$\pm$	0.09	&	7.11	$\pm$	0.10	&	42	&	7.11	$\pm$	0.13	&	13	&	1	\\
41484	&	71148	&	G5V	&	5838	$\pm$	39	&	4.58	$\pm$	0.12	&	1.30	$\pm$	0.33	&	0.01	$\pm$	0.05	&	7.51	$\pm$	0.06	&	57	&	7.52	$\pm$	0.07	&	12	&	1	\\
42074	&	72760	&	G5	&	5203	$\pm$	40	&	4.62	$\pm$	0.09	&	0.66	$\pm$	0.37	&	0.01	$\pm$	0.06	&	7.52	$\pm$	0.07	&	52	&	7.51	$\pm$	0.08	&	10	&	1	\\
42808	&	74576	&	K2V	&	4915	$\pm$	45	&	4.48	$\pm$	0.11	&	1.05	$\pm$	0.28	&	-0.16	$\pm$	0.09	&	7.35	$\pm$	0.11	&	51	&	7.34	$\pm$	0.14	&	9	&	5	\\
47592	&	84117	&	G0V	&	5991	$\pm$	62	&	4.21	$\pm$	0.17	&	1.89	$\pm$	0.36	&	-0.21	$\pm$	0.08	&	7.29	$\pm$	0.10	&	36	&	7.29	$\pm$	0.11	&	6	&	2	\\
48113	&	84737	&	G2V	&	5821	$\pm$	25	&	4.06	$\pm$	0.08	&	1.15	$\pm$	0.12	&	0.10	$\pm$	0.03	&	7.60	$\pm$	0.03	&	57	&	7.60	$\pm$	0.04	&	11	&	1	\\
51459	&	90839	&	F8V	&	6050	$\pm$	28	&	4.29	$\pm$	0.08	&	1.28	$\pm$	0.15	&	-0.15	$\pm$	0.04	&	7.35	$\pm$	0.04	&	62	&	7.35	$\pm$	0.05	&	13	&	4	\\
54745	&	97334	&	G0V	&	5865	$\pm$	45	&	4.46	$\pm$	0.13	&	1.48	$\pm$	0.34	&	-0.01	$\pm$	0.06	&	7.49	$\pm$	0.07	&	44	&	7.49	$\pm$	0.08	&	10	&	1	\\
56452	&	100623	&	K0V	&	5139	$\pm$	53	&	4.55	$\pm$	0.12	&	0.70	$\pm$	0.28	&	-0.38	$\pm$	0.07	&	7.12	$\pm$	0.08	&	58	&	7.12	$\pm$	0.11	&	10	&	4	\\
56997	&	101501	&	G8V	&	5591	$\pm$	49	&	4.60	$\pm$	0.14	&	0.93	$\pm$	0.20	&	0.01	$\pm$	0.04	&	7.50	$\pm$	0.04	&	63	&	7.51	$\pm$	0.06	&	13	&	4	\\
57757	&	102870	&	F8V	&	6044	$\pm$	40	&	4.02	$\pm$	0.12	&	1.41	$\pm$	0.18	&	0.09	$\pm$	0.04	&	7.59	$\pm$	0.05	&	60	&	7.59	$\pm$	0.06	&	12	&	4	\\
57939	&	103095	&	G8V	&	5144	$\pm$	77	&	4.05	$\pm$	0.20	&	0.77	$\pm$	0.49	&	-1.12	$\pm$	0.25	&	6.38	$\pm$	0.30	&	33	&	6.38	$\pm$	0.39	&	6	&	1	\\
61100	&	109011	&	K2V	&	4925	$\pm$	48	&	4.38	$\pm$	0.12	&	1.07	$\pm$	0.29	&	-0.34	$\pm$	0.11	&	7.16	$\pm$	0.12	&	36	&	7.16	$\pm$	0.17	&	8	&	1	\\
62523	&	111395	&	G7V	&	5677	$\pm$	33	&	4.65	$\pm$	0.08	&	0.64	$\pm$	0.20	&	0.22	$\pm$	0.03	&	7.72	$\pm$	0.04	&	58	&	7.72	$\pm$	0.03	&	12	&	1	\\
63742	&	113449	&	G5V	&	5050	$\pm$	55	&	4.51	$\pm$	0.13	&	0.96	$\pm$	0.28	&	-0.17	$\pm$	0.10	&	7.33	$\pm$	0.12	&	49	&	7.33	$\pm$	0.15	&	8	&	1	\\
64394	&	114710	&	G0V	&	6023	$\pm$	40	&	4.24	$\pm$	0.11	&	1.03	$\pm$	0.20	&	0.11	$\pm$	0.03	&	7.61	$\pm$	0.03	&	62	&	7.61	$\pm$	0.05	&	13	&	4	\\
64792	&	115383	&	G0Vs	&	6133	$\pm$	50	&	4.63	$\pm$	0.13	&	1.30	$\pm$	0.26	&	0.24	$\pm$	0.04	&	7.74	$\pm$	0.05	&	53	&	7.74	$\pm$	0.05	&	13	&	1	\\
65515	&	116956	&	G9IV-V	&	5214	$\pm$	34	&	4.65	$\pm$	0.08	&	0.93	$\pm$	0.31	&	0.03	$\pm$	0.05	&	7.53	$\pm$	0.07	&	49	&	7.53	$\pm$	0.07	&	8	&	1	\\
65530	&	117043	&	G6V	&	5610	$\pm$	38	&	4.38	$\pm$	0.11	&	0.73	$\pm$	0.17	&	0.29	$\pm$	0.03	&	7.79	$\pm$	0.04	&	58	&	7.79	$\pm$	0.03	&	13	&	1	\\
67620	&	120690	&	G5V	&	5720	$\pm$	33	&	4.63	$\pm$	0.09	&	0.60	$\pm$	0.39	&	0.05	$\pm$	0.04	&	7.55	$\pm$	0.06	&	49	&	7.54	$\pm$	0.05	&	11	&	3	\\
71743	&	128987	&	G6V	&	5511	$\pm$	123	&	4.86	$\pm$	0.29	&	0.80	$\pm$	0.48	&	0.01	$\pm$	0.11	&	7.51	$\pm$	0.12	&	48	&	7.51	$\pm$	0.16	&	5	&	2	\\
77408	&	141272	&	G8V	&	5191	$\pm$	31	&	4.64	$\pm$	0.07	&	0.87	$\pm$	0.21	&	-0.14	$\pm$	0.05	&	7.37	$\pm$	0.07	&	49	&	7.36	$\pm$	0.08	&	10	&	1	\\
77760	&	142373	&	F9V	&	5802	$\pm$	65	&	4.14	$\pm$	0.19	&	2.00	$\pm$	0.38	&	-0.50	$\pm$	0.07	&	7.00	$\pm$	0.09	&	50	&	7.00	$\pm$	0.11	&	13	&	1	\\
77801	&	142267	&	G0IV	&	5698	$\pm$	53	&	4.39	$\pm$	0.17	&	1.14	$\pm$	0.40	&	-0.50	$\pm$	0.10	&	7.00	$\pm$	0.13	&	42	&	7.00	$\pm$	0.14	&	9	&	3	\\
79672	&	146233	&	G1V	&	5804	$\pm$	81	&	4.43	$\pm$	0.20	&	1.01	$\pm$	0.24	&	0.04	$\pm$	0.06	&	7.54	$\pm$	0.09	&	59	&	7.54	$\pm$	0.09	&	12	&	4	\\
81300	&	149661	&	K2V	&	5192	$\pm$	57	&	4.65	$\pm$	0.16	&	0.33	$\pm$	0.46	&	0.05	$\pm$	0.06	&	7.55	$\pm$	0.07	&	55	&	7.55	$\pm$	0.10	&	11	&	4	\\
82588	&	152391	&	G8V	&	5442	$\pm$	51	&	4.62	$\pm$	0.16	&	0.34	$\pm$	0.43	&	0.04	$\pm$	0.05	&	7.54	$\pm$	0.05	&	41	&	7.54	$\pm$	0.07	&	13	&	1	\\
84862	&	157214	&	G0V	&	5663	$\pm$	34	&	4.40	$\pm$	0.12	&	1.32	$\pm$	0.34	&	-0.41	$\pm$	0.05	&	7.09	$\pm$	0.07	&	48	&	7.09	$\pm$	0.08	&	10	&	3	\\
91438	&	172051	&	G5V	&	5638	$\pm$	27	&	4.65	$\pm$	0.08	&	0.94	$\pm$	0.29	&	-0.21	$\pm$	0.04	&	7.29	$\pm$	0.06	&	60	&	7.29	$\pm$	0.06	&	13	&	5	\\
94346	&	180161	&	G8	&	5344	$\pm$	59	&	4.55	$\pm$	0.12	&	0.71	$\pm$	0.35	&	0.18	$\pm$	0.06	&	7.68	$\pm$	0.09	&	53	&	7.68	$\pm$	0.08	&	10	&	1	\\
96100	&	185144	&	K0V	&	5329	$\pm$	72	&	4.54	$\pm$	0.20	&	0.98	$\pm$	0.35	&	-0.26	$\pm$	0.12	&	7.24	$\pm$	0.13	&	60	&	7.24	$\pm$	0.18	&	13	&	4	\\
96895	&	186408	&	G2V	&	5760	$\pm$	74	&	4.35	$\pm$	0.24	&	1.05	$\pm$	0.18	&	0.02	$\pm$	0.06	&	7.52	$\pm$	0.08	&	53	&	7.52	$\pm$	0.10	&	10	&	3	\\
98819	&	190406	&	G1V	&	6067	$\pm$	32	&	4.64	$\pm$	0.08	&	1.20	$\pm$	0.22	&	0.12	$\pm$	0.03	&	7.62	$\pm$	0.05	&	41	&	7.62	$\pm$	0.04	&	5	&	2	\\
99240	&	190248	&	G5IV-V	&	5603	$\pm$	35	&	4.37	$\pm$	0.11	&	0.87	$\pm$	0.20	&	0.36	$\pm$	0.02	&	7.86	$\pm$	0.03	&	56	&	7.86	$\pm$	0.02	&	13	&	5	\\
105858	&	203608	&	F6V	&	5910	$\pm$	38	&	4.12	$\pm$	0.06	&	2.60	$\pm$	0.20	&	-0.84	$\pm$	0.04	&	6.66	$\pm$	0.02	&	42	&	6.66	$\pm$	0.07	&	13	&	5	\\
115331	&	220182	&	K1V	&	5455	$\pm$	48	&	4.65	$\pm$	0.15	&	0.72	$\pm$	0.26	&	0.11	$\pm$	0.04	&	7.61	$\pm$	0.05	&	39	&	7.61	$\pm$	0.07	&	12	&	1	\\
116613	&	222143	&	G5	&	5795	$\pm$	35	&	4.41	$\pm$	0.11	&	1.00	$\pm$	0.27	&	0.14	$\pm$	0.04	&	7.64	$\pm$	0.06	&	61	&	7.64	$\pm$	0.06	&	13	&	1	\\
116771	&	222368	&	F7V	&	6221	$\pm$	92	&	4.37	$\pm$	0.22	&	2.38	$\pm$	0.39	&	-0.13	$\pm$	0.04	&	7.37	$\pm$	0.10	&	55	&	7.37	$\pm$	0.12	&	13	&	4	\\
\noalign{\smallskip}\hline\noalign{\smallskip}
\multicolumn{12}{l}{$^{\dag}$Spectrograph: {\bf(1)} CAHA/FOCES; {\bf(2)} TNG/SARG; {\bf(3)} NOT/FIES; {\bf(4)} S$^{4}$N-McD {\bf(5)} 
S$^{4}$N-FEROS; {\bf(6)} ESO/FEROS ST-ECF Science Archive}\\
\hline
\end{longtable}
\end{tiny}
\end{center}
}


\end{document}